\documentclass[11pt]{article}
\usepackage{epsfig} 
\newcommand{\sect}[1]{ \section{#1} \setcounter{equation}{0} }

\newcommand{\Dslash}{D \! \! \! \! /}

\newcommand{\pslash}{p \! \! \! /}
\newcommand{\qslash}{q \! \! \! /}

\newcommand{\half}{\mbox{\small{$\frac{1}{2}$}}}

\newcommand{\muI}{\mu_{{\cal I}}}
\newcommand{\muIp}{\mu_{{\cal I}_+}}
\newcommand{\muIm}{\mu_{{\cal I}_-}}
\newcommand{\muIpm}{\mu_{{\cal I}_\pm}}
\newcommand{\muQ}{\mu_{{\cal Q}}}
\newcommand{\muQp}{\mu_{{\cal Q}_+}}
\newcommand{\muQm}{\mu_{{\cal Q}_-}}
\newcommand{\muR}{\mu_{{\cal R}}}
\newcommand{\muRp}{\mu_{{\cal R}_+}}
\newcommand{\muRm}{\mu_{{\cal R}_-}}
\newcommand{\muX}{\mu_{{\cal X}}}
\newcommand{\muXp}{\mu_{{\cal X}_+}}
\newcommand{\muXm}{\mu_{{\cal X}_-}}

\newcommand{\Nf}{N_{\!f}}

\newcommand{\NA}{N_{\!A}}

\begin{document}
\title{Power corrections to symmetric point vertices in Gribov-Zwanziger
theory}
\author{J.A. Gracey, \\ Theoretical Physics Division, \\ 
Department of Mathematical Sciences, \\ University of Liverpool, \\ P.O. Box 
147, \\ Liverpool, \\ L69 3BX, \\ United Kingdom.} 
\date{} 
\maketitle 

\vspace{5cm} 
\noindent 
{\bf Abstract.} The $3$-point vertices of QCD are examined at the symmetric
subtraction point at one loop in the Landau gauge in the presence of the Gribov
mass, $\gamma$. They are expanded in powers of $\gamma^2$ up to dimension four 
in order to determine the order of the leading correction. As well as analysing
the pure Gribov-Zwanziger Lagrangian, its extensions to include localizing 
ghost masses are also examined. For comparison a pure gluon mass term is also 
considered.

\vspace{-16cm}
\hspace{13.5cm}
{\bf LTH 959}

\newpage 

\sect{Introduction.}

In the late 1990's an interesting property of the running of an effective 
coupling constant derived from the triple gluon vertex in Quantum 
Chromodynamics (QCD) was revealed. In several articles, \cite{1,2,3,4,5}, using
lattice gauge theory techniques and restricting to the Landau gauge the 
effective coupling constant appeared to deviate from the expected behaviour in 
an energy range intermediate between high and low. Moreover the deviation 
between expected and measured behaviour could be fitted by a power law 
correction. While such discrepancies are not ordinarily unexpected in essence 
because of our knowledge of the operator product expansion, the power 
correction was claimed to correspond to a dimension {\em two} operator rather 
than a dimension four one, \cite{1,2,3,4,5}. The latter is usually associated 
with the vacuum expectation value of the square of the gluon field strength, 
which is gauge invariant, and is termed the gluon condensate. However, it was 
proposed that the operator associated with the dimension two correction was 
$\half {A^a_\mu}^2$ where $A^a_\mu$ is the gauge potential. Such a gauge 
variant operator is not excluded as ultimately the running coupling constant is
not a physical quantity. Subsequent to this there has been interest in trying 
to understand this dimension two operator both in the coupling constant 
situation and other quantities such as the gluon and Faddeev-Popov ghost 
propagators in the infrared. The condensation of a dimension two operator could
be related to confinement in the sense that it generates an effective gluon 
mass. See, for example, \cite{6} for a summary. Thus in the infrared the gluon 
propagator will freeze to a finite non-zero value at zero momentum. This is in 
accord with a long-standing result of Cornwall, \cite{7}, who established that 
the frozen gluon propagator follows from the dynamical generation of an
effective mass which is momentum dependent. In more recent years progress with
Landau gauge lattice computations has produced data which actually appears to 
support a frozen gluon propagator. For instance, see the early activity in this
respect \cite{8,9,10,11,12,13,14,15,16}. This should be qualified by the remark
that ensuring that one is in a properly fixed gauge free from the complications
of Gribov copies, \cite{17}, is a non-trivial exercise. Indeed it is perhaps 
fair to comment that this has yet to be fully resolved. Also the zero momentum 
regime is numerically difficult to achieve. Aside from the gluon propagator the
Faddeev-Popov ghost propagator has also been studied in the Landau gauge. It 
too can be used to study an effective running coupling constant at zero 
momentum via the ghost-gluon vertex which exploits properties derived from the 
Slavnov-Taylor identities, \cite{18}.

From the analytic point of view one can focus on the Landau gauge in the 
infrared. The seminal work in this area was by Gribov, \cite{17}, who
highlighted the inability to fix the gauge {\em globally} due to the presence
of Gribov copies. One property he established was that a less incorrectly fixed
Landau gauge could be effected with a modification of the Yang-Mills action. 
This restricted the path integral to the first Gribov region and introduced a 
non-local operator into the Lagrangian. Although a semi-classical approach was 
used in \cite{17} the non-locality was localized in a series of articles by 
Zwanziger, \cite{19,20,21,22,23,24,25,26,27}, to produce a local 
renormalizable Lagrangian. This construction meant that one could compute in 
the Gribov context and study infrared behaviour. Already noted in \cite{17} the
copies introduced a new mass called the Gribov mass, $\gamma$, which is not an 
independent parameter as it satisfies a gap equation. Its presence ensures that
the Faddeev-Popov ghost propagator enhances and the gluon propagator is 
suppressed in that it vanishes at zero momentum. While this does not accord 
with recent lattice data, modifications of the original Gribov-Zwanziger 
Lagrangian, \cite{28,29,30}, have been developed which do model the frozen 
gluon and non-enhanced Faddeev-Popov ghost propagator. Though such refinements 
are not unique. In \cite{30} it was demonstrated that several different 
localizing ghost condensations can model a frozen gluon. However each has a 
different prediction for other quantities which are not yet or as widely 
measured on the lattice. While there has been an intense amount of lattice 
activity on the propagators with as yet no resolution as to which refined 
solution is the leading candidate, the presence of an independent mass scale in
the pure or refined Gribov-Zwanziger Lagrangians could be the source of the 
power corrections of the effective coupling constants of \cite{1,2,3,4,5}. 
Therefore, it is the purpose of this article to explore this possibility and 
produce the one loop corrections to $3$-point vertices of QCD using the 
Gribov-Zwanziger setup. We will do this for the specific momentum configuration
of having the three external legs at non-zero momenta and with squared momenta 
all equal. This is known as the symmetric point and is a non-exceptional 
momentum configuration. Hence it will be free of infrared ambiguities which 
could plague the asymmetric subtraction point calculation which is an 
exceptional configuration. 

One advantage of this symmetric subtraction point is that by considering 
non-zero external momenta one will avoid having to go to the far infrared which
is beset with potential gauge fixing issues. In essence the momentum range we 
have in mind for comparisons to numerical work is the intermediate one where 
the explorations of \cite{1} were centred. In other words one is in a next to 
high energy approximation where one can formally access power corrections. At 
much lower energies the explicit expressions for vertex amplitudes would be 
complicated functions of the masses and momenta. These would then have to be 
expanded in a power series to reveal power corrections. This intermediate 
energy range is important from the Feynman diagram point of view. The main 
reason for this is that we can expand the Feynman graphs in a power series in a
mass scale. This avoids having to actually determine the explicit complicated 
functions and then expand them. Central to our calculations will be the use of 
the method of \cite{31} which allows us to correctly power expand Feynman 
integrals. Given that the Gribov mass is the associated scale in the pure 
Gribov-Zwanziger case there is no a priori reason to exclude it as being a 
potential source for the power corrections advanced in \cite{1,2,3,4,5}. For
instance, one could in principle relate $\gamma^2$ to the vacuum expectation 
value of $\half {A^a_\mu}^2$ precisely because $\gamma^2$ appears in the gluon 
propagator, \cite{17}. Hence $\langle \half {A^a_\mu}^2 \rangle$ will be 
proportional to $\gamma^2$ on dimensional grounds but both will be as 
indistinguishable from the other as the dimension two correction measured on 
the lattice. Thus in our exploration of this problem the aim will not only be 
the determination of whether there are dimension two, four or higher 
corrections but also to deduce the magnitude and sign of the coefficient. In 
studying the triple gluon, ghost-gluon and quark-gluon vertices we will be able
to devise a test which in principle could distinguish which of the refined 
solutions if any is favourable but crucially in a regime where deep infrared 
issues, such as Gribov copies, do not complicate a lattice study. However, as a
control on the Gribov mass investigation we will also repeat the same analysis 
for the case where the gluon has an explicit mass, $m$. Ordinarily such a naive
mass term breaks gauge invariance but one can have a non-local but gauge 
invariant gluon mass term. (See, for instance, \cite{32,33}.) This operator 
reduces to $\half m^2 {A^a_\mu}^2$ in the Landau gauge. Having such a control 
calculation to compare with the Gribov mass case is important as a gluon mass 
also mimicks a frozen gluon propagator. Thus that could actually be the source 
of the lattice computations. If that were the case then a symmetric vertex 
study could be used as a confirmatory test.

The paper is organized as follows. In section $2$ we discuss the general 
background to the Gribov-Zwanziger Lagrangian and its extensions which are 
needed for the computations we carry out. This includes the formalism we use to 
construct the three vertex functions at the symmetric point. Section $3$ 
focuses on general aspects of the calculation including how the Feynman 
integrals are deduced in a power series expansion using the technique devised
in \cite{31}. The subsequent four sections are devoted to recording the results
for the pure Gribov-Zwanziger case, what is termed the ${\cal Q}$ and 
${\cal R}$ solutions and a pure gluon mass respectively. We discuss our test in
the concluding section. Several appendices provide technical details which 
supplement the main discussion.

\sect{Background.}

We begin by recalling the main aspects of the Gribov-Zwanziger construction
which are necessary for our computations. In \cite{17} the Yang-Mills action
was modified in order to take into account the ambiguities which arise in the
gauge fixing procedure. With the restriction of the path integral to the first 
Gribov region the modification defines a boundary termed the Gribov horizon 
which manifests itself as a non-local dimension zero operator in the 
Lagrangian. It is related to the Faddeev-Popov operator, $\left( \partial^\mu 
D_\mu \right)^{ab}$, in that the boundary corresponds to the surface defined by
the first zeros of the operator. Thus the inverse operator will be infinite at 
the Gribov horizon and hence as long as there are no poles in the inverse one 
is within the Gribov region, \cite{17}. Originally in \cite{17}, the action was
treated in a semi-classical approximation in such a way that only the leading 
term of the Faddeev-Popov operator was used to define the boundary in the path 
integral. Subsequently, Zwanziger extended the analysis to all orders 
to produce the inverse Faddeev-Popov operator, \cite{21}. The resulting 
Lagrangian is, \cite{26}, 
\begin{equation}
L^{\mbox{\footnotesize{Gribov}}} ~=~ L^{\mbox{\footnotesize{QCD}}} ~+~
\frac{\gamma^4}{2} f^{eac} f^{ebd} A^a_\mu(x) \left( 
\frac{1}{\partial^\nu D_\nu} \right)^{cd} A^{b\,\mu}(x) ~-~ 
\frac{d \NA \gamma^4}{2g^2}
\label{laggrib}
\end{equation}
where $g$ is the coupling constant, $d$ is the spacetime dimension, $\NA$ is 
the dimension of the adjoint representation of the colour group whose structure
constants are $f^{abc}$. Here 
\begin{equation}
L^{\mbox{\footnotesize{QCD}}} ~=~ -~ \frac{1}{4} G_{\mu\nu}^a
G^{a \, \mu\nu} ~-~ \frac{1}{2\alpha} (\partial^\mu A^a_\mu)^2 ~-~
\bar{c}^a \partial^\mu D_\mu c^a ~+~ i \bar{\psi}^{iI} \Dslash \psi^{iI}
\end{equation}
is the usual QCD Lagrangian valid at high energy where $c^a$ is the 
Faddeev-Popov ghost and $\psi^{iI}$ are massless quarks. Although we have 
included the gauge parameter $\alpha$ associated with the linear covariant
gauge fixing we will perform all our calculations in the Landau gauge which is
$\alpha$~$=$~$0$. In (\ref{laggrib}) the mass parameter $\gamma$ is known as 
the Gribov mass and is not an independent parameter as it satifies a gap 
equation derived from the horizon, (\ref{gribhor}). Only when $\gamma$ 
satisfies the gap equation is one actually in the gauge theory \cite{17,21,26}.
The field independent term of (\ref{laggrib}) ensures the non-triviality of the 
condition since, \cite{26}, 
\begin{equation}
f^{eac} f^{ebd} \left\langle A^a_\mu(x) \left( \frac{1}{\partial^\nu D_\nu}
\right)^{cd} A^{b\,\mu}(x) \right\rangle ~=~ \frac{d N_A}{g^2} ~.
\label{gribhor}
\end{equation}
The presence of the non-local operator implies that the gluon propagator is
modified from that which is used at high energy. Specifically, \cite{17}, 
\begin{equation}
\langle A^a_\mu(p) A^b_\nu(-p) \rangle ~=~ -~ 
\frac{\delta^{ab}p^2}{[(p^2)^2+C_A\gamma^4]} P_{\mu\nu}(p) 
\end{equation}
which behaves as 
$-$~$\frac{P_{\mu\nu}(p)}{p^2} \delta^{ab}$ at large momenta 
where 
\begin{equation}
P_{\mu\nu}(p) ~=~ \eta_{\mu\nu} ~-~ \frac{p_\mu p_\nu}{p^2} ~.
\end{equation}
At low momenta the propagator tends to zero which is the gluon suppression,
\cite{17}.

While (\ref{laggrib}) extends the Yang-Mills action to incorporate the copy
issue, from a practical point of view the non-local term means that one cannot
use it for explicit calculations. To circumvent this Zwanziger managed to
localize the non-locality in several articles, \cite{20,21,26}, to produce a 
local Lagrangian. Its renormalizability was established in \cite{26,34,35}. In 
order to achieve the localization localizing ghost fields were introduced,
\cite{26}, so that (\ref{laggrib}) is replaced by
\begin{eqnarray}
L^{\mbox{\footnotesize{GZ}}} &=& L^{\mbox{\footnotesize{QCD}}} ~+~
\frac{1}{2} \rho^{ab \, \mu} \partial^\nu \left( D_\nu \rho_\mu
\right)^{ab} ~+~ \frac{i}{2} \rho^{ab \, \mu} \partial^\nu
\left( D_\nu \xi_\mu \right)^{ab} ~-~ \frac{i}{2} \xi^{ab \, \mu}
\partial^\nu \left( D_\nu \rho_\mu \right)^{ab} \nonumber \\
&& +~ \frac{1}{2} \xi^{ab \, \mu} \partial^\nu \left( D_\nu \xi_\mu
\right)^{ab} ~-~ \bar{\omega}^{ab \, \mu} \partial^\nu \left( D_\nu \omega_\mu
\right)^{ab} ~-~ \frac{1}{\sqrt{2}} g f^{abc} \partial^\nu
\bar{\omega}^{ae}_\mu \left( D_\nu c \right)^b \rho^{ec \, \mu} \nonumber \\
&& -~ \frac{i}{\sqrt{2}} g f^{abc} \partial^\nu \bar{\omega}^{ae}_\mu
\left( D_\nu c \right)^b \xi^{ec \, \mu} ~-~ i \gamma^2 f^{abc} A^{a \, \mu}
\xi^{bc}_\mu ~-~ \frac{d \NA \gamma^4}{2g^2} ~.
\label{laggz}
\end{eqnarray}
Here $\rho^{ab}_\mu$ and $\xi^{ab}_\mu$ are real fields and we have chosen this
version over the complex fields of the earlier localization, \cite{20,21,26}. 
Accompanying these bosonic fields are the other localizing ghosts, 
$\omega^{ab}_\mu$ and $\bar{\omega}^{ab}_\mu$, which are Grassmann. They are 
required to ensure that the ultraviolet structure of the theory such as 
asymptotic freedom is not upset nor renormalizability lost by the sole presence
of the bosonic ghosts. The other main feature is that the dimension zero 
operator of (\ref{laggrib}) is translated into a dimension two mass-like term 
which mixes $A^a_\mu$ and $\xi^{ab}_\mu$. In effect this term corresponds to 
the non-local term which can be clearly seen by recalling the relation
\begin{equation}
A^a_\mu ~=~ -~ \frac{i}{C_A \gamma^2} f^{abc} \left( \partial^\nu D_\nu \xi_\mu
\right)^{bc}
\end{equation}
which is deduced from the $\xi^{ab}_\mu$ equation of motion. The original
horizon condition (\ref{gribhor}) becomes 
\begin{equation}
f^{abc} \left\langle A^{a\,\mu} (x) \xi^{bc}_\mu(x) \right\rangle ~=~
\frac{i d\NA \gamma^2}{g^2}
\end{equation}
in (\ref{laggz}). With the extra fields there is an extended set of propagators
which are, \cite{26}, 
\begin{eqnarray}
\langle A^a_\mu(p) A^b_\nu(-p) \rangle &=& -~ 
\frac{\delta^{ab}p^2}{[(p^2)^2+C_A\gamma^4]} P_{\mu\nu}(p) ~~~,~~~
\langle A^a_\mu(p) \xi^{bc}_\nu(-p) \rangle ~=~ 
\frac{i f^{abc}\gamma^2}{[(p^2)^2+C_A\gamma^4]} P_{\mu\nu}(p) \nonumber \\
\langle \xi^{ab}_\mu(p) \xi^{cd}_\nu(-p) \rangle &=& -~ 
\frac{\delta^{ac}\delta^{bd}}{p^2}\eta_{\mu\nu} ~+~
\frac{f^{abe}f^{cde}\gamma^4}{p^2[(p^2)^2+C_A\gamma^4]} P_{\mu\nu}(p) ~~,~~ 
\langle A^a_\mu(p) \rho^{bc}_\nu(-p) \rangle ~=~ 0 \nonumber \\ 
\langle \rho^{ab}_\mu(p) \rho^{cd}_\nu(-p) \rangle &=& 
\langle \omega^{ab}_\mu(p) \bar{\omega}^{cd}_\nu(-p) \rangle ~=~ -~ 
\frac{\delta^{ac}\delta^{bd}}{p^2} \eta_{\mu\nu} ~~,~~  
\langle \xi^{ab}_\mu(p) \rho^{cd}_\nu(-p) \rangle ~=~ 0 ~. 
\label{propgz}
\end{eqnarray} 
The suppressed gluon propagator emerges again. Despite the presence of the 
mixed propagator it is possible to compute one and two loop corrections to 
various quantities such as the gap equation for $\gamma$, \cite{17,26,36}.

While this was the standard Lagrangian used to incorporate Gribov copies it did
not cover the observed behaviour on the lattice since the gluon propagator does
not freeze to zero but to a finite non-zero value. To account for this in the 
Gribov-Zwanziger context various extensions to (\ref{laggz}) have been
considered in \cite{28,29,30}. Each generalization can be summarized in the 
addition of a mass operator for the localizing ghost sector. The most general 
such operator is, \cite{30}, 
\begin{equation}
{\cal O} ~=~ \left[ \mu_{{\cal Q}}^2 \delta^{ac} \delta^{bd} ~+~ 
\mu_{{\cal W}}^2 f^{ace} f^{bde} ~+~ \frac{\mu_{{\cal R}}^2}{C_A} 
f^{abe} f^{cde} ~+~ \mu_{{\cal S}}^2 d_A^{abcd} ~+~ 
\frac{\mu_{{\cal P}}^2}{\NA} \delta^{ab} \delta^{cd} ~+~ \mu_{{\cal T}}^2 
\delta^{ad} \delta^{bc} \right] {\cal O}^{abcd}
\label{opgen}
\end{equation}
where 
\begin{equation}
{\cal O}^{abcd} ~=~ \frac{1}{2} \left[ \rho^{ab} \rho^{cd} ~+~
i \xi^{ab} \rho^{cd} ~-~ i \rho^{ab} \xi^{cd} ~+~ \xi^{ab} \xi^{cd} \right] ~-~
\bar{\omega}^{ab} \omega^{cd} ~.
\end{equation}
Each colour structure is tagged with a mass, $\muI$, where we use the same
labels and conventions as were used in \cite{30}. The tensor $d_A^{abcd}$ is
totally symmetric and is defined by, \cite{37}, 
\begin{equation}
d_A^{abcd} ~=~ \frac{1}{6} \mbox{Tr} \left( T_A^a T_A^{(b} T_A^c T_A^{d)}
\right) ~.
\end{equation}
Overall the additional operator is BRST invariant and satisfies a 
Slavnov-Taylor identity so that the operator renormalization constant is 
related to the $\xi^{ab}_\mu$ wave function renormalization constant, 
\cite{28,29,30}. With this additional term the propagators (\ref{propgz}) 
become significantly more complicated. These were analysed at length in 
\cite{30} where the full set for $SU(3)$ were recorded explicitly. In general 
and for certain specific cases the extra mass parameters $\muI$ lead to gluon 
propagators which freeze to a non-zero finite value. In \cite{29} one of these 
cases was examined which in the notation of (\ref{opgen}) was termed the 
${\cal Q}$ solution. However, in \cite{30} it was noted that this solution was 
not unique and that another specific single mass parameter solution could be 
equally viable. This was the ${\cal R}$ case. It was argued in \cite{30} that 
this was a more natural solution than the ${\cal Q}$ case since a non-zero 
$\muR$ would correspond to a condensation of the operator $\xi^{ab}_\mu 
\xi^{ab\,\mu}$ consistent with the structure of the pure Gribov-Zwanziger 
propagators, \cite{30}. To see the frozen gluon propagator at the outset the 
propagators for these two specific solutions are 
\begin{eqnarray}
\langle A^a_\mu(p) A^b_\nu(-p) \rangle_{{\cal Q}} &=& -~ 
\frac{\delta^{ab}[p^2+\mu_{{\cal Q}}^2]}
{[(p^2)^2+\mu_{{\cal Q}}^2p^2+C_A\gamma^4]} P_{\mu\nu}(p) ~~,~~ 
\langle A^a_\mu(p) \rho^{bc}_\nu(-p) \rangle_{{\cal Q}} ~=~ 0 \nonumber \\
\langle A^a_\mu(p) \xi^{bc}_\nu(-p) \rangle_{{\cal Q}} &=& 
\frac{i f^{abc}\gamma^2}{[(p^2)^2+\mu_{{\cal Q}}^2p^2+C_A\gamma^4]} 
P_{\mu\nu}(p) ~~,~~ 
\langle \xi^{ab}_\mu(p) \rho^{cd}_\nu(-p) \rangle_{{\cal Q}} ~=~ 0 \nonumber \\ 
\langle \xi^{ab}_\mu(p) \xi^{cd}_\nu(-p) \rangle_{{\cal Q}} &=& -~ 
\frac{\delta^{ac}\delta^{bd}}{[p^2+\mu_{{\cal Q}}^2]}\eta_{\mu\nu} ~+~
\frac{f^{abe}f^{cde}\gamma^4}
{[p^2+\mu_{{\cal Q}}^2][(p^2)^2+\mu_{{\cal Q}}^2p^2+C_A\gamma^4]} 
P_{\mu\nu}(p) \nonumber \\ 
\langle \rho^{ab}_\mu(p) \rho^{cd}_\nu(-p) \rangle_{{\cal Q}} &=& 
\langle \omega^{ab}_\mu(p) \bar{\omega}^{cd}_\nu(-p) \rangle_{{\cal Q}} ~=~ -~ 
\frac{\delta^{ac}\delta^{bd}}{[p^2+\mu_{{\cal Q}}^2]} \eta_{\mu\nu} 
\end{eqnarray} 
and
\begin{eqnarray}
\langle A^a_\mu(p) A^b_\nu(-p) \rangle_{{\cal R}} &=& -~ 
\frac{\delta^{ab}[p^2+\mu_{{\cal R}}^2]}
{[(p^2)^2+\mu_{{\cal R}}^2p^2+C_A\gamma^4]} P_{\mu\nu}(p) ~~,~~ 
\langle A^a_\mu(p) \rho^{bc}_\nu(-p) \rangle_{{\cal R}} ~=~ 0 \nonumber \\
\langle A^a_\mu(p) \xi^{bc}_\nu(-p) \rangle_{{\cal R}} &=& 
\frac{i f^{abc}\gamma^2}{[(p^2)^2+\mu_{{\cal R}}^2p^2+C_A\gamma^4]} 
P_{\mu\nu}(p) ~~,~~ 
\langle \xi^{ab}_\mu(p) \rho^{cd}_\nu(-p) \rangle_{{\cal R}} ~=~ 0 \nonumber \\ 
\langle \xi^{ab}_\mu(p) \xi^{cd}_\nu(-p) \rangle_{{\cal R}} &=& -~ 
\frac{\delta^{ac}\delta^{bd}}{p^2}\eta_{\mu\nu} ~+~
\frac{f^{abe}f^{cde}[\mu_{{\cal R}}^2p^2+C_A\gamma^4]}
{C_Ap^2[(p^2)^2+\mu_{{\cal R}}^2p^2+C_A\gamma^4]} P_{\mu\nu}(p) \nonumber \\
&& +~ \frac{f^{abe}f^{cde}\mu_{{\cal R}}^2}{C_Ap^2[p^2+\mu_{{\cal R}}^2]} 
L_{\mu\nu}(p) \nonumber \\ 
\langle \rho^{ab}_\mu(p) \rho^{cd}_\nu(-p) \rangle_{{\cal R}} &=& 
\langle \omega^{ab}_\mu(p) \bar{\omega}^{cd}_\nu(-p) \rangle_{{\cal R}} ~=~ -~ 
\frac{\delta^{ac}\delta^{bd}}{p^2} \eta_{\mu\nu} ~+~ 
\frac{f^{abe}f^{cde}\mu_{{\cal R}}^2}{C_Ap^2[p^2+\mu_{{\cal R}}^2]} 
\eta_{\mu\nu} ~.
\end{eqnarray} 
It is these propagators which we will use as part of our study of the power 
corrections to the $3$-point vertices. While the gluon sectors of each are 
formally equivalent the key differences are in the localizing ghost 
propagators. In the ${\cal Q}$ case there are no massless factors in any of the
propagators whereas there are massless modes in the ${\cal R}$ solution. In
\cite{30} it was noted that this led to different infrared properties of the 
localizing ghost propagators after the gap equation for $\gamma$ is satisfied. 
Such differing behaviour can be used to distinguish from these solutions if 
lattice data was available for the localizing ghost propagators. 

We now turn to the formalism relating to specific Green's functions we will 
compute in a power series expansion. These are 
\begin{eqnarray}
\left. \left\langle A^a_\mu(p) A^b_\nu(q) A^c_\sigma(r)
\right\rangle \right|_{p^2 = q^2 = - \mu^2} &=& f^{abc}
\left. \Sigma^{\mbox{\footnotesize{ggg}}}_{\mu \nu \sigma}(p,q,\gamma^2,\muI^2)
\right|_{p^2 = q^2 = - \mu^2} \nonumber \\
\left. \left\langle c^a(p) \bar{c}^b(q) A^c_\sigma(r)
\right\rangle \right|_{p^2 = q^2 = - \mu^2} &=& f^{abc}
\left. \Sigma^{\mbox{\footnotesize{ccg}}}_\sigma(p,q,\gamma^2,\muI^2)
\right|_{p^2 = q^2 = - \mu^2} \nonumber \\ 
\left. \left\langle \psi^i(p) \bar{\psi}^j(q) A^c_\sigma(r)
\right\rangle \right|_{p^2 = q^2 = - \mu^2} &=& T^c_{ij}
\left. \Sigma^{\mbox{\footnotesize{qqg}}}_\sigma(p,q,\gamma^2,\muI^2)
\right|_{p^2 = q^2 = - \mu^2} 
\label{gfamps}
\end{eqnarray}
where
\begin{equation}
r ~=~ -~ p ~-~ q
\end{equation}
and each Green's function corresponds respectively to the triple gluon, 
ghost-gluon and quark-gluon vertices. The independent external momenta flowing 
in through two of the external legs are $p$ and $q$ and neither are nullified 
so that we are in a non-exceptional momentum configuration which does not 
suffer from infrared issues. All squared momenta are held at the same value
\begin{equation}
p^2 ~=~ q^2 ~=~ r^2 ~=~ -~ \mu^2
\end{equation}
which implies
\begin{equation}
pq ~=~ \frac{1}{2} \mu^2
\end{equation}
where $\mu$ is the mass scale introduced to ensure that the coupling constant
remains dimensionless in $d$-dimensional spacetime as we will be using 
dimensional regularization throughout. Our regularizing parameter will be
$\epsilon$ where $d$~$=$~$4$~$-$~$2\epsilon$. As each Green's function carries
colour and Lorentz indices we have to decompose them into scalar amplitudes.
For the former we have done this in (\ref{gfamps}). This is relatively
straightforward since to the loop order we are working there is only one colour
tensor for each vertex function which is evident from the explicit
computations. For the Lorentz sector we have to introduce a set of basis
tensors which are built from $\eta_{\mu\nu}$ and $p_\mu$ and $q_\mu$. Thus
\begin{eqnarray}
\left. \frac{}{} \Sigma^{\mbox{\footnotesize{ggg}}}_{\mu \nu \sigma}
(p,q,\gamma^2,\muI^2) \right|_{p^2 = q^2 = - \mu^2} &=& \sum_{k=1}^{14}
{\cal P}^{\mbox{\footnotesize{ggg}}}_{(k) \, \mu \nu \sigma }(p,q) \,
\Sigma^{\mbox{\footnotesize{ggg}}}_{(k)}(p,q,\gamma^2,\muI^2) \nonumber \\
\left. \frac{}{} \Sigma^{\mbox{\footnotesize{ccg}}}_\sigma(p,q,\gamma^2,\muI^2)
\right|_{p^2 = q^2 = - \mu^2} &=& \sum_{k=1}^{2}
{\cal P}^{\mbox{\footnotesize{ccg}}}_{(k) \, \sigma }(p,q) \,
\Sigma^{\mbox{\footnotesize{ccg}}}_{(k)}(p,q,\gamma^2,\muI^2) \nonumber \\
\left. \frac{}{} \Sigma^{\mbox{\footnotesize{qqg}}}_\sigma(p,q,\gamma^2,\muI^2)
\right|_{p^2 = q^2 = - \mu^2} &=& \sum_{k=1}^{6}
{\cal P}^{\mbox{\footnotesize{qqg}}}_{(k) \, \sigma }(p,q) \,
\Sigma^{\mbox{\footnotesize{qqg}}}_{(k)}(p,q,\gamma^2,\muI^2) 
\end{eqnarray}
which defines the scalar amplitudes for each vertex function. The explicit 
tensors for each case are given in Appendix A and we use the same set as was
used in \cite{38}. We note that away from the symmetric point restriction the 
basis will involve more tensors. Also the basis is not unique and there are 
other choices. It turns out that for the triple gluon vertex from the explicit
computations of \cite{38} we have checked that to two loops one can write the 
vertex function more compactly in terms of {\em three} tensors. One of these, 
for instance, corresponds to the Feynman rule of the vertex itself when 
$r$~$=$~$-$~$p$~$-$~$q$. The other two do not involve $\eta_{\mu\nu}$ and their
explicit forms are given in Appendix A. Therefore, here we choose to work in 
this more compact basis and replace the first equation of (\ref{gfamps}) by 
\begin{equation}
\left. \frac{}{} \tilde{\Sigma}^{\mbox{\footnotesize{ggg}}}_{\mu \nu \sigma}
(p,q,\gamma^2,\muI^2) \right|_{p^2 = q^2 = - \mu^2} ~=~ \sum_{k=1}^{3}
\tilde{\cal P}^{\mbox{\footnotesize{ggg}}}_{(k) \, \mu \nu \sigma }(p,q) \,
\tilde{\Sigma}^{\mbox{\footnotesize{ggg}}}_{(k)}(p,q,\gamma^2,\muI^2) ~. 
\end{equation}
To distinguish this basis from the previous set the amplitudes will have a
tilde. In order to calculate each scalar amplitude we use the same projection
approach as \cite{38}. Briefly each Green's function is multiplied by a linear
combination of the basis tensors in $d$-dimensions. The coefficients are found 
by first constructing the matrix of products of all the basis tensors. Then the
inverse of this matrix gives the linear combination. Denoting this inverse 
matrix by ${\cal M}$ we have 
\begin{eqnarray}
f^{abc} \tilde{\Sigma}^{\mbox{\footnotesize{ggg}}}_{(k)}(p,q,\gamma^2,\muI^2) 
&=& \tilde{\cal M}^{\mbox{\footnotesize{ggg}}}_{kl} \left(
\tilde{\cal P}^{\mbox{\footnotesize{ggg}} \, \mu \nu \sigma}_{(l)}(p,q) \left.
\left\langle A^a_\mu(p) A^b_\nu(q) A^c_\sigma(r)
\right\rangle \right )\right|_{p^2 = q^2 = - \mu^2} \nonumber \\
f^{abc} \Sigma^{\mbox{\footnotesize{ccg}}}_{(k)}(p,q,\gamma^2,\muI^2) &=&
{\cal M}^{\mbox{\footnotesize{ccg}}}_{kl} \left(
{\cal P}^{\mbox{\footnotesize{ccg}} \, \sigma}_{(l)}(p,q) \left.
\left\langle c^a(p) \bar{c}^b(q) A^c_\sigma(r)
\right\rangle \right) \right|_{p^2 = q^2 = - \mu^2} \nonumber \\
T^c_{ij} \Sigma^{\mbox{\footnotesize{qqg}}}_{(k)}(p,q,\gamma^2,\muI^2) &=&
{\cal M}^{\mbox{\footnotesize{qqg}}}_{kl} \left(
{\cal P}^{\mbox{\footnotesize{qqg}} \, \sigma}_{(l)}(p,q) \left.
\left\langle \psi^i(p) \bar{\psi}^j(q) A^c_\sigma(r)
\right\rangle \right) \right|_{p^2 = q^2 = - \mu^2}
\end{eqnarray}
where $k$ and $l$ are the matrix labels. The explicit matrices are given in
Appendix A. For the quark sector we have the additional issue of spinor indices
to account for in the amplitude decomposition. So as well as building the 
tensor basis from $\eta_{\mu\nu}$, $p_\mu$ and $q_\mu$ one has also to include 
$\gamma_\mu$. This implies that products of $\gamma$-matrices can arise in the 
tensor basis and since we will be working in $d$-dimensions it is natural to 
use the generalized basis of $\gamma$-matrices which spans spinor space in 
$d$-dimensions, \cite{39,40,41}. These are denoted by $\Gamma_{(n)}^{\mu_1 
\ldots \mu_n}$ where $n$ is a positive integer and defined by
\begin{equation}
\Gamma_{(n)}^{\mu_1 \ldots \mu_n} ~=~ \gamma^{[\mu_1} \ldots \gamma^{\mu_n]}
\end{equation}
where $1/n!$ is included in the antisymmetrization. The algebra of these has
been studied at length in \cite{41,42,43}. Though for constructing the matrix
${\cal M}^{\mbox{\footnotesize{qqg}}}_{kl}$ a useful property is, \cite{42,44},
\begin{equation}
\mbox{tr} \left( \Gamma_{(m)}^{\mu_1 \ldots \mu_m}
\Gamma_{(n)}^{\nu_1 \ldots \nu_n} \right) ~ \propto ~ \delta_{mn}
I^{\mu_1 \ldots \mu_m \nu_1 \ldots \nu_n}
\end{equation}
which partitions the matrix where $I^{\mu_1 \ldots \mu_m \nu_1 \ldots \nu_n}$
is the unit matrix on the generalized spinor space.

\sect{Calculational method.}

We now detail the overall method we have used to determine the power 
corrections to the one loop vertices at the symmetric point in each of the 
various Lagrangians we are interested in. We have followed a general approach
which allows us to construct routines for all possible mass configurations. For
instance, examining the propagators all possible one loop $3$-point functions
will have at most three different non-zero mass scales. However, since we are
at a symmetric point and there are massless poles in the propagator sets, there
are actually only seven distinct basic mass distribution configuration we have 
to consider. If we define
\begin{equation}
I(m_1^2,m_2^2,m_3^2) ~=~ \left. \int \frac{d^dk}{(2\pi)^d} \, 
\frac{1}{[k^2+m_1^2][(k-p)^2+m_2^2][(k+q)^2+m_3^2]} \right|_{p^2=q^2=-\mu^2}
\label{idef}
\end{equation}
then the basic master integral structures from the point of view of mass
distribution across the propagators are $I(0,0,0)$, $I(m_1^2,0,0)$,
$I(m_1^2,m_1^2,0)$, $I(m_1^2,m_2^2,0)$, $I(m_1^2,m_1^2,m_1^2)$, 
$I(m_1^2,m_1^2,m_2^2)$ and $I(m_1^2,m_2^2,m_3^2)$. Here none of the $m_i$ are 
equal. We note that (\ref{idef}) is the basic structure since any of the Gribov
or Stingl type propagators can always be written as the product of two
canonical propagators and then partial fractioned. The fact that we are at a 
symmetric subtraction point means that $I(m_1^2,m_2^2,m_3^2)$ is totally 
symmetric in its arguments. We have not included powers of the propagators in 
the definition of $I(m_1^2,m_2^2,m_3^2)$ as we use the standard approach of 
breaking up all the contributing Feynman graphs into scalar integrals. These 
are determined after the general projection method. Then numerator scalar 
products are rewritten in terms of the propagator factors. For one loop 
$3$-point functions at the symmetric point there are no irreducible numerators.
To proceed further we have to write these integrals, which may have propagators
to a negative power or positive power greater than unity, in terms of the basic
master integrals for each of the seven mass distributions. For the massless 
case we did this in \cite{38} at one {\em and} two loops. For the massive cases
we extend that approach which used the Laporta algorithm \cite{44}. That 
method, \cite{44}, uses integration by parts to establish algebraic relations 
between all the integrals which arise. As there is an overredundancy in the 
relations, one can relate all integrals down to a basic set known as masters. 
The values for these are determined by other methods and hence the whole 
calculation is complete. In terms of tools we have used the {\sc Reduze} 
implementation, \cite{45}, which uses the {\sc GiNaC} symbolic manipulation 
system, \cite{46}, and is written in C++. The {\sc Reduze} package creates a 
database for each topology and mass distribution and one extracts the relations
which are needed for the specific computation. Aside from the massless one used
in \cite{38} we have constructed five other one loop databases. That for 
$I(m_1^2,m_2^2,m_3^2)$ is not needed as no integral of this mass distribution 
arises for propagator powers other than unity. Once the databases are 
determined the required integral relations are written in {\sc Form}, 
\cite{47}, and included in the automatic computation routine. To generate the 
Feynman graphs for each of the three vertices we use {\sc Qgraf}, \cite{48}. 
For the Gribov-Zwanziger case there are $30$ one loop graphs and for the other 
two vertices there are $3$ one loop graphs in both cases. 

Having broken down the computation of the amplitudes of each vertex into the
basic master integrals we have to substitute the explicit values for each. In
general terms at this stage there are three basic configurations which 
correspond to having one, two or three propagators. By this we mean that in the
latter two cases any two or three of the propagators present in (\ref{idef}).
For the one and two propagator cases for different masses these integrals are
known exactly and their power series expansion can then be substituted after
carrying out a Taylor expansion. For $I(m_1^2,m_2^2,m_3^2)$ itself the explicit 
result is not known for any combination of non-zero $m_i$ except when all three
masses are zero. Therefore, since we are only interested in the power series
expansion we follow the method of \cite{31}. This method allows one to expand
Feynman integrals in powers of $m^2/\mu^2$ where $m$ is a generic mass scale
deriving from a propagator in the original integral and $\mu$ is our common
scale here for the squared momentum of the external legs. While \cite{31}
detailed at length the expansion of the two loop self-energy topology as an
example the method is general and we use the general formalism that was
provided there. If in general we denote by $J$ one of our master integrals
$I(m_1^2,m_2^2,m_3^2)$ with at least one $m_i$~$\neq$~$0$ then the asymptotic
expansion is, \cite{31},
\begin{equation}
J_\Gamma ~ \sim ~ \sum_\lambda \, J_{\Gamma/\lambda} \circ
{\cal T}_{\{m_i\};\{q_i\}} J_\lambda ~.
\end{equation}
We use similar notation as \cite{31} and note that $\Gamma$ is the original
Feynman diagram, $\lambda$ are subgraphs which arise in the asymptotic
expansion and $m_i$ formally represent the masses on each propagator of
$\Gamma$. (In \cite{31} $\gamma$ was used for the subgraphs but we use 
$\lambda$ here to avoid confusion with the Gribov mass which is a parameter in 
the actual expansion.) These are necessary to counteract the infrared 
infinities which arise if one naively expands the original integral in powers 
of $1/\mu^2$. That is always the first term in the expansion and in that case 
the subgraph $\lambda$ is the unit graph. The other non-unit graphs in the sum 
are constructed from all possible routings of the external momenta around the 
graph. In the two loop example detailed in \cite{31} there was only one such
momenta unlike the two here. However, it is straightforward to see that there
are three such graphs since with two momenta there are three ways to route the 
external momenta. In other words in each of the three graphs one of the three 
propagators of (\ref{idef}) will have no external momenta. For each of these 
three cases the subgraph $\lambda$ is expanded in powers of the masses $m_i$ 
and the momenta $q_i$ external to that subgraph itself. The terms of this 
Taylor expansion, denoted by ${\cal T}_{\{m_i\};\{q_i\}} J_\lambda$, are then 
substituted in the reduced diagram $J_{\Gamma/\lambda}$ before performing the 
integration over the loop momentum, \cite{31}. For the expansion the leading 
term will involve massless $3$-point integrals which can be reduced to the one 
loop massless master of \cite{49} using the {\sc Reduze} database we have 
already constructed. The remaining three terms reduce to one loop massive 
tadpoles which are readily evaluated. In Appendix C we have given various 
examples of the expansion of master integrals which were used within our 
calculations. 

\sect{Power corrections in the Gribov-Zwanziger Lagrangian.}

We now turn to the mundane task of recording the explicit results for each of
the three vertices for various versions of the Gribov-Zwanziger Lagrangian,
its extensions and the gluon mass case\footnote{Electronic forms of all the
results are contained in an attached data file.}. In this section we focus on
the pure Gribov-Zwanziger Lagrangian, (\ref{laggz}), where there are no mass 
terms deriving from localizing ghost field masses. Throughout this and 
subsequent sections we give the symmetric vertices to and including the term 
which corresponds to the dimension four correction. While we could in principle
include higher order corrections, there does not appear to be any practical 
reason to do this at present since such terms would be difficult to extract 
numerically from the lattice. First, we record that the triple gluon vertex 
structure for 
(\ref{laggz}) is 
\begin{eqnarray}
\tilde{\Sigma}^{ggg}_{(1)}(p,q,\gamma^2,0) &=& 
\tilde{\Sigma}^{ggg}_{(1)}(p,q,0,0) \nonumber \\
&& +~ \left[ \left[ \frac{13}{6} + \frac{7\pi^2}{36} 
- \frac{7}{24} \psi^\prime \left( \frac{1}{3} \right) 
- \frac{1}{2} \ln \left[ \frac{C_A\gamma^4}{\mu^4} \right] \right] 
\frac{C_A^2 \gamma^4}{\mu^4} + O \left( \frac{\gamma^6}{\mu^6} \right) 
\right] a \nonumber \\
&& +~ O(a^2) \nonumber \\
\tilde{\Sigma}^{ggg}_{(2)}(p,q,\gamma^2,0) &=& 
\tilde{\Sigma}^{ggg}_{(2)}(p,q,0,0) \nonumber \\
&& +~ \left[ \frac{3\pi}{32} \frac{C_A^{3/2} \gamma^2}{\mu^2} 
+ \left[ \frac{289}{144} + \frac{787}{192} \ln \left[ 
\frac{C_A\gamma^4}{\mu^4} \right] \right] \frac{C_A^2 \gamma^4}{\mu^4} ~+~ 
O \left( \frac{\gamma^6}{\mu^6} \right) \right] a ~+~ O(a^2) \nonumber \\
\tilde{\Sigma}^{ggg}_{(3)}(p,q,\gamma^2,0) &=& 
\tilde{\Sigma}^{ggg}_{(3)}(p,q,0,0) \nonumber \\
&& + \left[ \frac{3\pi}{32} \frac{C_A^{3/2} \gamma^2}{\mu^2}
+ \left[ \frac{25}{18} \psi^\prime \left( \frac{1}{3} \right) 
- \frac{25}{27} \pi^2 - \frac{2033}{576} + \frac{799}{192} \ln 
\left[ \frac{C_A\gamma^4}{\mu^4} \right] \right] \frac{C_A^2 \gamma^4}{\mu^4}
\right. \nonumber \\
&& \left. ~~~~+~ O \left( \frac{\gamma^6}{\mu^6} \right) 
\right] a ~+~ O(a^2) ~. 
\end{eqnarray}
Interestingly there is no dimension two correction for channel $1$ which is the
channel corresponding to the original triple gluon vertex Feynman rule. The 
other amplitudes have a dimension two correction and moreover, that term is the
same in both cases. The explicit values of the $\gamma^2$~$=$~$0$ 
amplitudes for this and the other cases are given in Appendix B. For the 
ghost-gluon and quark-gluon vertices we have  
\begin{eqnarray}
\Sigma^{ccg}_{(1)}(p,q,\gamma^2,0) &=& \Sigma^{ccg}_{(1)}(p,q,0,0)
\nonumber \\
&& - \left[ \frac{9\pi}{32} \frac{C_A^{3/2} \gamma^2}{\mu^2}
+ \left[ \frac{377}{2304} + \frac{1}{144} \psi^\prime \left( \frac{1}{3} 
\right) - \frac{1}{216} \pi^2 - \frac{1}{12} \ln \left[ 
\frac{C_A\gamma^4}{\mu^4} \right] \right] \frac{C_A^2 \gamma^4}{\mu^4} \right. 
\nonumber \\
&& \left. ~~~~+~ O \left( \frac{\gamma^6}{\mu^6} \right) \right] a ~+~ O(a^2) 
\nonumber \\ 
\Sigma^{ccg}_{(2)}(p,q,\gamma^2,0) &=& \Sigma^{ccg}_{(2)}(p,q,0,0)
\nonumber \\
&& + \left[ \frac{3\pi}{16} \frac{C_A^{3/2} \gamma^2}{\mu^2}
+ \left[ \frac{1}{144} \psi^\prime \left( \frac{1}{3} \right) 
- \frac{1}{216} \pi^2 - \frac{7}{2304} - \frac{1}{48} \ln \left[ 
\frac{C_A\gamma^4}{\mu^4} \right] \right] \frac{C_A^2 \gamma^4}{\mu^4} \right. 
\nonumber \\
&& \left. ~~~~+~ O \left( \frac{\gamma^6}{\mu^6} \right) \right] a ~+~ O(a^2)
\end{eqnarray}
and
\begin{eqnarray}
\Sigma^{qqg}_{(1)}(p,q,\gamma^2,0) &=& \Sigma^{qqg}_{(1)}(p,q,0,0)
\nonumber \\
&& + \left[ \left[ \frac{3}{8} C_F + \frac{3}{16} C_A \right] 
\frac{\pi\sqrt{C_A}\gamma^2}{\mu^2} \right. \nonumber \\
&& \left. ~~~~+ 
\left[ \left[ \frac{67}{36} - \frac{4}{9} \psi^\prime \left( \frac{1}{3} 
\right) + \frac{8}{27} \pi^2 - \frac{2}{3} \ln \left[ 
\frac{C_A\gamma^4}{\mu^4} \right] \right] C_F \right. \right. \nonumber \\
&& \left. \left. ~~~~~~~+
\left[ \frac{11}{18} \psi^\prime \left( \frac{1}{3} \right)
- \frac{2941}{1152} - \frac{11}{27} \pi^2 + \frac{31}{48} \ln \left[ 
\frac{C_A\gamma^4}{\mu^4} \right] \right] C_A \right] 
\frac{C_A \gamma^4}{\mu^4} \right. \nonumber \\
&& \left. ~~~~+~ O \left( \frac{\gamma^6}{\mu^6} \right) \right] a ~+~ O(a^2) 
\nonumber \\ 
\Sigma^{qqg}_{(2)}(p,q,\gamma^2,0) &=& \Sigma^{qqg}_{(5)}(p,q,\gamma^2,0) 
\nonumber \\ 
&=& \Sigma^{qqg}_{(2)}(p,q,0,0) \nonumber \\
&& + \left[ -~ \frac{3\pi}{16} C_A \frac{\sqrt{C_A}\gamma^2}{\mu^2}
\right. \nonumber \\
&& \left. ~~~~+
\left[ \left[ \frac{43}{18} - \frac{8}{9} \psi^\prime \left( \frac{1}{3} 
\right) + \frac{16}{27} \pi^2 - \frac{5}{6} \ln \left[ 
\frac{C_A\gamma^4}{\mu^4} \right] \right] C_F \right. \right. \nonumber \\
&& \left. \left. ~~~~~~~+
\left[ \frac{14}{9} \psi^\prime \left( \frac{1}{3} \right)
- \frac{1351}{288} - \frac{28}{27} \pi^2 + \frac{325}{192} \ln \left[ 
\frac{C_A\gamma^4}{\mu^4} \right] \right] C_A \right] 
\frac{C_A \gamma^4}{\mu^4} \right. \nonumber \\
&& \left. ~~~~+~ O \left( \frac{\gamma^6}{\mu^6} \right) \right] a ~+~ O(a^2) 
\nonumber \\ 
\Sigma^{qqg}_{(3)}(p,q,\gamma^2,0) &=& \Sigma^{qqg}_{(4)}(p,q,\gamma^2,0) 
\nonumber \\ 
&=& \Sigma^{qqg}_{(3)}(p,q,0,0) \nonumber \\
&& + \left[ \left[ \frac{3}{16} C_A - \frac{3}{4} C_F \right] 
\frac{\pi\sqrt{C_A}\gamma^2}{\mu^2} \right. \nonumber \\
&& \left. ~~~~+ 
\left[ \left[ \frac{1}{2} \ln \left[ \frac{C_A\gamma^4}{\mu^4} \right]
- \frac{4}{3} \right] C_F \right. \right. \nonumber \\
&& \left. \left. ~~~~~~~~+
\left[ \frac{2}{3} \psi^\prime \left( \frac{1}{3} \right)
- \frac{917}{576} - \frac{4}{9} \pi^2 + \frac{109}{192} \ln \left[ 
\frac{C_A\gamma^4}{\mu^4} \right] \right] C_A \right] 
\frac{C_A \gamma^4}{\mu^4} \right. \nonumber \\
&& \left. ~~~~+~ O \left( \frac{\gamma^6}{\mu^6} \right) \right] a ~+~ O(a^2) 
\nonumber \\ 
\Sigma^{qqg}_{(6)}(p,q,\gamma^2,0) &=& \Sigma^{qqg}_{(6)}(p,q,0,0) 
\nonumber \\
&& + \left[ \left[ \frac{3}{4} C_F - \frac{9}{8} C_A \right] 
\frac{\pi\sqrt{C_A}\gamma^2}{\mu^2} \right. \nonumber \\
&& \left. ~~~~+ 
\left[ \left[ \frac{31}{18} - \frac{1}{3} \ln \left[ \frac{C_A\gamma^4}{\mu^4} 
\right] \right] C_F 
+ \left[ \frac{1}{24} \ln \left[ \frac{C_A\gamma^4}{\mu^4} \right]
- \frac{991}{576} \right] C_A \right] \frac{C_A \gamma^4}{\mu^4} \right. 
\nonumber \\
&& \left. ~~~~+~ O \left( \frac{\gamma^6}{\mu^6} \right) 
\right] a ~+~ O(a^2) ~. 
\end{eqnarray}
For the latter the equivalences between various amplitudes merely reflects the
underlying left-right symmetry of the vertex in the choice of basis tensors we 
have used. We have not imposed this symmetry within the computation but instead
it has emerged naturally and is regarded as a minor internal check on the 
programming. Unlike the triple gluon case there is a dimension two correction 
for all channels of both these vertices.

\sect{${\cal Q}$ solution.} 

Next we turn to what is termed the {\cal Q} solution in the notation used in
\cite{29}. In this and subsequent sections the results are more involved due to
the presence of an additional mass scale corresponding to the localizing ghost 
field mass of the particular solution. Therefore, we will have various
combinations of $\gamma$ and $\muI$. To compactify notation we will define
\begin{equation}
\muIpm^2 ~=~ \frac{1}{2} \left[ \muI^2 ~\pm~ \sqrt{[\muI^4-4C_A\gamma^4]}
\right] 
\end{equation}
where ${\cal I}$ corresponds to the particular solution of interest. For the
triple gluon vertex we have the expressions
\begin{eqnarray}
\tilde{\Sigma}^{ggg}_{(1)}(p,q,\gamma^2,\muQ^2) &=& 
\tilde{\Sigma}^{ggg}_{(1)}(p,q,0,0) \nonumber \\
&& +~ \left[ \left[ \frac{11}{64} \muQ^2 \sqrt{[\muQ^4-4C_A\gamma^4]} 
\ln \left[ \frac{\muQp^2}{\muQm^2} \right] 
+ \frac{3\muQ^4}{32} \ln \left[ \frac{C_A\gamma^4}{\muQ^4} \right] 
\right. \right. \nonumber \\
&& \left. \left. ~~~~~~- 
\frac{5\muQ^6}{64\sqrt{[\muQ^4-4C_A\gamma^4]}}
\ln \left[ \frac{\muQp^2}{\muQm^2} \right] 
\right. \right. \nonumber \\
&& \left. \left. ~~~~~~+ 
\left[ \frac{13}{6} + \frac{7\pi^2}{36} - \frac{7}{24} \psi^\prime 
\left( \frac{1}{3} \right) - \frac{1}{2} \ln \left[ \frac{C_A\gamma^4}{\mu^4} 
\right] \right] C_A\gamma^4 \right] \frac{C_A}{\mu^4} \right. \nonumber \\
&& \left. ~~~~~~+ O \left( \frac{\muQ^6}{\mu^6} \right) \right] a ~+~ O(a^2) 
\nonumber \\ 
\tilde{\Sigma}^{ggg}_{(2)}(p,q,\gamma^2,\muQ^2) &=& 
\tilde{\Sigma}^{ggg}_{(2)}(p,q,0,0) \nonumber \\
&& + \left[ \left[ \frac{3\muQ^4}{64\sqrt{[\muQ^4-4C_A\gamma^4]}}
\ln \left[ \frac{\muQp^2}{\muQm^2} \right] 
- \frac{3}{64} \sqrt{[\muQ^4-4C_A\gamma^4]}
\ln \left[ \frac{\muQp^2}{\muQm^2} \right] 
\right] \frac{C_A}{\mu^2}
\right. \nonumber \\ 
&& \left. ~~~~~~+ \left[ -~ \frac{1075}{768} \muQ^2 
\sqrt{[\muQ^4-4C_A\gamma^4]} \ln \left[ \frac{\muQp^2}{\muQm^2} \right] 
- \frac{3\muQ^4}{4} \ln \left[ \frac{C_A\gamma^4}{\muQ^4} \right] 
\right. \right. \nonumber \\
&& \left. \left. ~~~~~~~~~~~+ 
\frac{499\muQ^6}{768\sqrt{[\muQ^4-4C_A\gamma^4]}}
\ln \left[ \frac{\muQp^2}{\muQm^2} \right] 
\right. \right. \nonumber \\
&& \left. \left. ~~~~~~~~~~~+ 
\left[ \frac{289}{144} 
+ \frac{787}{192} \ln \left[ \frac{C_A\gamma^4}{\mu^4} 
\right] \right] C_A\gamma^4 \right] \frac{C_A}{\mu^4} 
+ O \left( \frac{\muQ^6}{\mu^6} \right) \right] a ~+~ O(a^2) \nonumber \\ 
\tilde{\Sigma}^{ggg}_{(3)}(p,q,\gamma^2,\muQ^2) &=& 
\tilde{\Sigma}^{ggg}_{(3)}(p,q,0,0) \nonumber \\
&& + \left[ \left[ \frac{3\muQ^4}{64\sqrt{[\muQ^4-4C_A\gamma^4]}}
\ln \left[ \frac{\muQp^2}{\muQm^2} \right] 
- \frac{3}{64} \sqrt{[\muQ^4-4C_A\gamma^4]}
\ln \left[ \frac{\muQp^2}{\muQm^2} \right] \right] \frac{C_A}{\mu^2}
\right. \nonumber \\ 
&& \left. ~~~~~~+ \left[ -~ \frac{1087}{768} \muQ^2 
\sqrt{[\muQ^4-4C_A\gamma^4]} \ln \left[ \frac{\muQp^2}{\muQm^2} \right] 
- \frac{3\muQ^4}{4} \ln \left[ \frac{C_A\gamma^4}{\muQ^4} \right] 
\right. \right. \nonumber \\
&& \left. \left. ~~~~~~~~~~+ 
\frac{511C_A\muQ^6}{768\sqrt{[\muQ^4-4C_A\gamma^4]}}
\ln \left[ \frac{\muQp^2}{\muQm^2} \right] 
\right. \right. \nonumber \\
&& \left. \left. ~~~~~~~~~~+ 
\left[ \frac{25}{18} \psi^\prime \left( \frac{1}{3} \right) - \frac{2033}{576} 
- \frac{25\pi^2}{27} 
+ \frac{799}{192} \ln \left[ \frac{C_A\gamma^4}{\mu^4} 
\right] \right] C_A\gamma^4 \right] \frac{C_A}{\mu^4} 
\right. \nonumber \\
&& \left. ~~~~~~+ O \left( \frac{\muQ^6}{\mu^6} \right) \right] a ~+~ O(a^2) 
\end{eqnarray}
where like the pure Gribov-Zwanziger case the structure of dimension two 
corrections are the same. This is somewhat unexpected as the presence of the
extra mass does not induce a lower order correction. Therefore, if one were
using the absence of a dimension two correction in lattice data in channel $1$,
if for example that were the case, then one could not conclude that the pure
Gribov-Zwanziger case is an explanation since ${\cal Q}$ has the same 
qualitative feature. We note that here and subsequently we have used $O( 
\muI^6/\mu^6)$ to indicate we are dropping dimension six terms and have used 
$\muI^6$ within the order symbol to indicate a generic dimension six mass 
scale. In practice this could be a dependent on $\gamma$ too but it would 
complicate the notation. For the other two vertices we have
\begin{eqnarray}
\Sigma^{ccg}_{(1)}(p,q,\gamma^2,\muQ^2) &=&
\Sigma^{ccg}_{(1)}(p,q,0,0) \nonumber \\
&& + \left[ \left[ -~ \frac{9C_A \gamma^4}{16\sqrt{[\muQ^4-4C_A\gamma^4]}}
\ln \left[ \frac{\muQp^2}{\muQm^2} \right] \right] \frac{C_A}{\mu^2}
\right. \nonumber \\
&& \left. ~~~~+
\left[ \frac{C_A \gamma^4\muQ^2}{12\sqrt{[\muQ^4-4C_A\gamma^4]}}
\ln \left[ \frac{\muQp^2}{\muQm^2} \right]
\right. \right. \nonumber \\
&& \left. \left. ~~~~+
\left[ \frac{\pi^2}{216} - \frac{1}{144} \psi^\prime \left( \frac{1}{3} \right)
- \frac{377}{2304} + \frac{1}{12} \ln \left[ \frac{C_A\gamma^4}{\mu^4}
\right] \right] C_A\gamma^4 \right] \frac{C_A}{\mu^4}
\right. \nonumber \\
&& \left. ~~~~+ O \left( \frac{\muQ^6}{\mu^6} \right) \right] a ~+~ O(a^2)
\nonumber \\
\Sigma^{ccg}_{(2)}(p,q,\gamma^2,\muQ^2) &=&
\Sigma^{ccg}_{(2)}(p,q,0,0) \nonumber \\
&& + \left[ \left[ \frac{3C_A \gamma^4}{8\sqrt{[\muQ^4-4C_A\gamma^4]}}
\ln \left[ \frac{\muQp^2}{\muQm^2} \right] \right] \frac{C_A}{\mu^2}
\right. \nonumber \\
&& \left. ~~~~+
\left[ -~ \frac{C_A \gamma^4\muQ^2}{48\sqrt{[\muQ^4-4C_A\gamma^4]}}
\ln \left[ \frac{\muQp^2}{\muQm^2} \right]
\right. \right. \nonumber \\
&& \left. \left. ~~~~+
\left[ \frac{1}{144} \psi^\prime \left( \frac{1}{3} \right)
- \frac{\pi^2}{216}
- \frac{7}{2304} - \frac{1}{48} \ln \left[ \frac{C_A\gamma^4}{\mu^4}
\right] \right] C_A\gamma^4 \right] \frac{C_A}{\mu^4}
\right. \nonumber \\
&& \left. ~~~~+ O \left( \frac{\muQ^6}{\mu^6} \right) \right] a ~+~ O(a^2)
\end{eqnarray}
and 
\begin{eqnarray}
\Sigma^{qqg}_{(1)}(p,q,\gamma^2,\muQ^2) &=& 
\Sigma^{qqg}_{(1)}(p,q,0,0) \nonumber \\
&& + \left[ \left[ \left[ \frac{3}{4} C_F + \frac{3}{8} C_A \right] 
\frac{C_A\gamma^4}{\sqrt{[\muQ^4-4C_A\gamma^4]}} 
\ln \left[ \frac{\muQp^2}{\muQm^2} \right] \right] \frac{1}{\mu^2} 
\right. \nonumber \\
&& \left. ~~~+
\left[ \left[  
- \frac{2C_A\muQ^2\gamma^4}{3\sqrt{[\muQ^4-4C_A\gamma^4]}}
\ln \left[ \frac{\muQp^2}{\muQm^2} \right] 
\right. \right. \right. \nonumber \\
&& \left. \left. \left. ~~~~~~+
\left[ \frac{67}{36} + \frac{8\pi^2}{27} - \frac{4}{9} \psi^\prime 
\left( \frac{1}{3} \right) - \frac{2}{3} \ln \left[ \frac{C_A\gamma^4}{\mu^4} 
\right] \right] C_A\gamma^4 \right] C_F \right. \right. \nonumber \\
&& \left. \left. ~~~+
\left[ \frac{31C_A\muQ^2\gamma^4}{48\sqrt{[\muQ^4-4C_A\gamma^4]}}
\ln \left[ \frac{\muQp^2}{\muQm^2} \right] 
\right. \right. \right. \nonumber \\
&& \left. \left. \left. ~~~~~~+
\left[ \frac{11}{18} \psi^\prime \left( \frac{1}{3} \right) 
- \frac{2941}{1152} - \frac{11\pi^2}{27} 
+ \frac{31}{48} \ln \left[ \frac{C_A\gamma^4}{\mu^4} \right] 
\right] C_A\gamma^4 \right] C_A \right] \frac{1}{\mu^4} \right. \nonumber \\
&& \left. +~ O \left( \frac{\muQ^6}{\mu^6} \right) \right] a ~+~ O(a^2) 
\nonumber \\ 
\Sigma^{qqg}_{(2)}(p,q,\gamma^2,\muQ^2) &=& 
\Sigma^{qqg}_{(5)}(p,q,\gamma^2,\muQ^2) \nonumber \\
&=& \Sigma^{qqg}_{(2)}(p,q,0,0) \nonumber \\
&& + \left[ \left[ -~ \frac{3C_A^2\gamma^4}{8\sqrt{[\muQ^4-4C_A\gamma^4]}} 
\ln \left[ \frac{\muQp^2}{\muQm^2} \right] \right] \frac{1}{\mu^2} 
\right. \nonumber \\
&& \left. ~~~+
\left[ \left[  
-~ \frac{5C_A\muQ^2\gamma^4}{6\sqrt{[\muQ^4-4C_A\gamma^4]}}
\ln \left[ \frac{\muQp^2}{\muQm^2} \right] 
\right. \right. \right. \nonumber \\
&& \left. \left. \left. ~~~~~~~~~~+
\left[ \frac{43}{18} + \frac{16\pi^2}{27} - \frac{8}{9} \psi^\prime 
\left( \frac{1}{3} \right) - \frac{5}{6} \ln \left[ \frac{C_A\gamma^4}{\mu^4} 
\right] \right] C_A\gamma^4 \right] C_F
\right. \right. \nonumber \\
&& \left. \left. ~~~~~~~+
\left[ \left[ \frac{14}{9} \psi^\prime \left( \frac{1}{3} \right) 
- \frac{1351}{288} - \frac{28\pi^2}{27} 
+ \frac{325}{192} \ln \left[ \frac{C_A\gamma^4}{\mu^4} \right] 
\right] C_A\gamma^4 \right. \right. \right. \nonumber \\ 
&& \left. \left. \left. ~~~~~~~~~~~+
\frac{325C_A\muQ^2\gamma^4}{192\sqrt{[\muQ^4-4C_A\gamma^4]}}
\ln \left[ \frac{\muQp^2}{\muQm^2} \right] \right] C_A \right] \frac{1}{\mu^4} 
+ O \left( \frac{\muQ^6}{\mu^6} \right) \right] a \nonumber \\
&& +~ O(a^2) \nonumber \\ 
\Sigma^{qqg}_{(3)}(p,q,\gamma^2,\muQ^2) &=& 
\Sigma^{qqg}_{(4)}(p,q,\gamma^2,\muQ^2) \nonumber \\
&=& \Sigma^{qqg}_{(3)}(p,q,0,0) \nonumber \\
&& + \left[ \left[ \left[ \frac{3}{8} C_A - \frac{3}{2} C_F \right] 
\frac{C_A\gamma^4}{\sqrt{[\muQ^4-4C_A\gamma^4]}} 
\ln \left[ \frac{\muQp^2}{\muQm^2} \right] \right] \frac{1}{\mu^2} 
\right. \nonumber \\
&& \left. ~~~+
\left[ \left[ \frac{C_A\muQ^2\gamma^4}{2\sqrt{[\muQ^4-4C_A\gamma^4]}}
\ln \left[ \frac{\muQp^2}{\muQm^2} \right] 
+ \left[ \frac{1}{2} \ln \left[ \frac{C_A\gamma^4}{\mu^4} 
\right] - \frac{4}{3} \right] C_A\gamma^4 \right] C_F 
\right. \right. \nonumber \\
&& \left. \left. ~~~~~~~+
\left[ \frac{109C_A\muQ^2\gamma^4}{192\sqrt{[\muQ^4-4C_A\gamma^4]}}
\ln \left[ \frac{\muQp^2}{\muQm^2} \right] 
\right. \right. \right. \nonumber \\
&& \left. \left. \left. ~~~~~~~~~~~+
\left[ \frac{2}{3} \psi^\prime \left( \frac{1}{3} \right) 
- \frac{917}{576} - \frac{4\pi^2}{9} 
+ \frac{109}{192} \ln \left[ \frac{C_A\gamma^4}{\mu^4} \right] 
\right] C_A\gamma^4 \right] C_A \right] \frac{1}{\mu^4} 
\right. \nonumber \\
&& \left. ~~~+ O \left( \frac{\muQ^6}{\mu^6} \right) \right] a ~+~ O(a^2) 
\nonumber \\ 
\Sigma^{qqg}_{(6)}(p,q,\gamma^2,\muQ^2) &=& 
\Sigma^{qqg}_{(6)}(p,q,0,0) \nonumber \\
&& + \left[ \left[ \left[ \frac{3}{2} C_F - \frac{9}{4} C_A \right] 
\frac{C_A\gamma^4}{\sqrt{[\muQ^4-4C_A\gamma^4]}} 
\ln \left[ \frac{\muQp^2}{\muQm^2} \right] \right] \frac{1}{\mu^2} 
\right. \nonumber \\
&& \left. ~~~+
\left[ \left[  
\left[ \frac{31}{18} - \frac{1}{3} \ln \left[ \frac{C_A\gamma^4}{\mu^4} 
\right] \right] C_A\gamma^4 
- \frac{C_A\muQ^2\gamma^4}{3\sqrt{[\muQ^4-4C_A\gamma^4]}}
\ln \left[ \frac{\muQp^2}{\muQm^2} \right] \right] C_F
\right. \right. \nonumber \\
&& \left. \left. ~~~+
\left[ \frac{C_A\muQ^2\gamma^4}{24\sqrt{[\muQ^4-4C_A\gamma^4]}}
\ln \left[ \frac{\muQp^2}{\muQm^2} \right] 
\right. \right. \right. \nonumber \\
&& \left. \left. \left. ~~~~~~~+
\left[ \frac{1}{24} \ln \left[ \frac{C_A\gamma^4}{\mu^4} \right] 
- \frac{991}{576} \right]
C_A\gamma^4 \right] C_A \right] \frac{1}{\mu^4} 
\right. \nonumber \\
&& \left. ~~~~+ O \left( \frac{\muQ^6}{\mu^6} \right) \right] a ~+~ O(a^2) ~. 
\end{eqnarray} 
Again these expressions all have dimension two corrections albeit complicated.
However, they have the same qualitative structure as the pure Gribov-Zwanziger
case. In the above expressions for all the amplitudes we have checked that the
results of the previous section are reproduced as $\muQ^2$~$\rightarrow$~$0$.
This also applies to results in subsequent sections in the appropriate
$\muI^2$~$\rightarrow$~$0$ limit. We have also checked that in the 
$\gamma^2$~$\rightarrow$~$0$ limit there are no corrections at all as there
should be since then there is no horizon condition. 

\sect{${\cal R}$ solution.} 

Structurally, the expressions for the ${\cal R}$ are very similar to those for
${\cal Q}$. For instance, in many cases the differences are only in the 
numerical coefficients. Though the mass is $\muR$ rather than $\muQ$ of course. 
Therefore, we will make minimal comment on these parallel results. For the 
triple gluon vertex we have,
\begin{eqnarray}
\tilde{\Sigma}^{ggg}_{(1)}(p,q,\gamma^2,\muR^2) &=& 
\tilde{\Sigma}^{ggg}_{(1)}(p,q,0,0) \nonumber \\
&& +~ \left[ \left[ \frac{3}{32} \muR^2 \sqrt{[\muR^4-4C_A\gamma^4]} 
\ln \left[ \frac{\muRp^2}{\muRm^2} \right] 
+ \frac{3\muR^4}{32} \ln \left[ \frac{C_A\gamma^4}{\muR^4} \right] 
\right. \right. \nonumber \\
&& \left. \left. ~~~~- 
\frac{5C_A\muR^2\gamma^4}{16\sqrt{[\muR^4-4C_A\gamma^4]}}
\ln \left[ \frac{\muRp^2}{\muRm^2} \right] 
\right. \right. \nonumber \\
&& \left. \left. ~~~~+ 
\left[ \frac{13}{6} + \frac{7\pi^2}{36} - \frac{7}{24} \psi^\prime 
\left( \frac{1}{3} \right) - \frac{1}{2} \ln \left[ \frac{C_A\gamma^4}{\mu^4} 
\right] \right] C_A\gamma^4 \right] \frac{C_A}{\mu^4} \right. \nonumber \\
&& \left. ~~~~+ O \left( \frac{\muR^6}{\mu^6} \right) \right] a ~+~ O(a^2) 
\nonumber \\ 
\tilde{\Sigma}^{ggg}_{(2)}(p,q,\gamma^2,\muR^2) &=& 
\tilde{\Sigma}^{ggg}_{(2)}(p,q,0,0) \nonumber \\
&& + \left[ \left[ \frac{3C_A\gamma^4}{16\sqrt{[\muR^4-4C_A\gamma^4]}}
\ln \left[ \frac{\muRp^2}{\muRm^2} \right] \right] \frac{C_A}{\mu^2}
\right. \nonumber \\ 
&& \left. ~~~~+ \left[ -~ \frac{3}{4} \muR^2 
\sqrt{[\muR^4-4C_A\gamma^4]} \ln \left[ \frac{\muRp^2}{\muRm^2} \right] 
- \frac{3\muR^4}{4} \ln \left[ \frac{C_A\gamma^4}{\muR^4} \right] 
\right. \right. \nonumber \\
&& \left. \left. ~~~~~~~~~+ 
\frac{499C_A\muR^2\gamma^4}{192\sqrt{[\muR^4-4C_A\gamma^4]}}
\ln \left[ \frac{\muRp^2}{\muRm^2} \right] 
\right. \right. \nonumber \\
&& \left. \left. ~~~~~~~~~+ 
\left[ \frac{289}{144} 
+ \frac{787}{192} \ln \left[ \frac{C_A\gamma^4}{\mu^4} 
\right] \right] C_A\gamma^4 \right] \frac{C_A}{\mu^4} 
+ O \left( \frac{\muR^6}{\mu^6} \right) \right] a ~+~ O(a^2) \nonumber \\ 
\tilde{\Sigma}^{ggg}_{(3)}(p,q,\gamma^2,\muR^2) &=& 
\tilde{\Sigma}^{ggg}_{(3)}(p,q,0,0) \nonumber \\
&& + \left[ \left[ \frac{3C_A\gamma^4}{16\sqrt{[\muR^4-4C_A\gamma^4]}}
\ln \left[ \frac{\muRp^2}{\muRm^2} \right] \right] \frac{C_A}{\mu^2}
\right. \nonumber \\ 
&& \left. ~~~~+ \left[ -~ \frac{3}{4} \muR^2 
\sqrt{[\muR^4-4C_A\gamma^4]} \ln \left[ \frac{\muRp^2}{\muRm^2} \right] 
- \frac{3\muR^4}{4} \ln \left[ \frac{C_A\gamma^4}{\muR^4} \right] 
\right. \right. \nonumber \\
&& \left. \left. ~~~~~~~~~+ 
\frac{511C_A\muR^2\gamma^4}{192\sqrt{[\muR^4-4C_A\gamma^4]}}
\ln \left[ \frac{\muRp^2}{\muRm^2} \right] 
\right. \right. \nonumber \\
&& \left. \left. ~~~~~~~~~+ 
\left[ \frac{25}{18} \psi^\prime \left( \frac{1}{3} \right) - \frac{2033}{576} 
- \frac{25\pi^2}{27} 
+ \frac{799}{192} \ln \left[ \frac{C_A\gamma^4}{\mu^4} 
\right] \right] C_A\gamma^4 \right] \frac{C_A}{\mu^4} 
\right. \nonumber \\
&& \left. ~~~~+ O \left( \frac{\muR^6}{\mu^6} \right) \right] a ~+~ O(a^2) ~. 
\end{eqnarray}
Again there is no dimension two correction for channel $1$. For the other two
vertices it transpires that aside from mapping the masses 
$\muQ$~$\leftrightarrow$~$\muR$ the expressions for the ghost-gluon and 
quark-gluon vertices are formally the same at one loop. This is not unexpected 
if one considers the contributing Feynman diagrams. The propagators for 
${\cal Q}$ and ${\cal R}$ differ only in the localizing ghost sector. Since the 
gluon propagator is the only propagator from the gauge sector which contributes
to the ghost-gluon and quark-gluon vertices at one loop and is formally the 
same for both solutions then the respective vertex functions have to be the 
same. 

\sect{Explicit gluon mass.}

So far we have concentrated on Gribov and Stingl type propagators for the gluon
sector of Yang-Mills based on the pure and refined Gribov-Zwanziger
Lagrangians. However, an alternative scenario is that a frozen gluon
propagator in the infrared could be as a result of a pure gluon mass. While
such a term ordinarily breaks gauge invariance, it is possible to construct a 
gauge invariant mass operator for the gluon. However, such an operator has to 
be non-local, \cite{32,33}, and could be associated with an analogous ghost
operator which is non-local. The origin for the latter observation is that it
is possible to have a BRST invariant gluon mass term which was considered in
\cite{50}. There in order to make that local gluon mass operator BRST invariant
one had to include a ghost mass term which was dependent on the gauge 
parameter. In either case when one restricts both to the Landau gauge the 
non-locality disappears in the first case and in the second only one of the two
terms contributing to the BRST operator survives. In both cases in the Landau 
gauge
\begin{equation}
{\cal O}_{A_\mu^2} ~=~ \frac{\mu_{{\cal X}}^2}{2} A^a_\mu A^{a\,\mu} 
\label{glmass}
\end{equation}
emerges as the mass term for the gluon. We use the same notation as \cite{30}
for consistency and will regard the presence of (\ref{glmass}) in a Lagrangian 
as our control calculation. With (\ref{glmass}) we can derive the propagator 
for $A^a_\mu$. However, in order to compare with the previous two cases we will
also add (\ref{glmass}) into the Gribov-Zwanziger Lagrangian. Thus our 
propagators are 
\begin{eqnarray}
\langle A^a_\mu(p) A^b_\nu(-p) \rangle_{{\cal X}} &=& -~ 
\frac{\delta^{ab}p^2}{[(p^2)^2+\mu_{{\cal X}}^2p^2+C_A\gamma^4]} P_{\mu\nu}(p) 
\nonumber \\
\langle A^a_\mu(p) \xi^{bc}_\nu(-p) \rangle_{{\cal X}} &=& 
\frac{i f^{abc}\gamma^2}{[(p^2)^2+\mu_{{\cal X}}^2p^2+C_A\gamma^4]} 
P_{\mu\nu}(p) \nonumber \\
\langle A^a_\mu(p) \rho^{bc}_\nu(-p) \rangle_{{\cal X}} &=& 0 ~~,~~ 
\langle \xi^{ab}_\mu(p) \rho^{cd}_\nu(-p) \rangle_{{\cal X}} ~=~ 0 \nonumber \\ 
\langle \xi^{ab}_\mu(p) \xi^{cd}_\nu(-p) \rangle_{{\cal X}} &=& -~ 
\frac{\delta^{ac}\delta^{bd}}{p^2}\eta_{\mu\nu} ~+~
\frac{f^{abe}f^{cde}\gamma^4}{p^2[(p^2)^2+\mu_{{\cal X}}^2p^2+C_A\gamma^4]} 
P_{\mu\nu}(p) \nonumber \\ 
\langle \rho^{ab}_\mu(p) \rho^{cd}_\nu(-p) \rangle_{{\cal X}} &=& 
\langle \omega^{ab}_\mu(p) \bar{\omega}^{cd}_\nu(-p) \rangle_{{\cal X}} ~=~ -~ 
\frac{\delta^{ac}\delta^{bd}}{p^2} \eta_{\mu\nu} ~. 
\end{eqnarray} 
Clearly in the limit where $\gamma$~$\rightarrow$~$0$ the first term reduces to
what one expects for a gluon propagator with a naive mass term. Although the
$\rho^{ab}_\mu$, $\xi^{ab}_\mu$ and $\omega^{ab}_\mu$ propagators remain their
contribution cancels within any calculation as if they were not present in the
first case in this limit. Though for any results we present we have checked
this directly by using the usual Yang-Mills Lagrangian without including a
Gribov mass term.

The procedure to compute the vertex functions for this case is precisely the
same as the previous two sections. However, we first give the results for a
pure mass term, $\muX$, in the absence of $\gamma$. For the triple gluon
vertex we have 
\begin{eqnarray}
\tilde{\Sigma}^{ggg}_{(1)}(p,q,0,\muX^2) &=& 
\tilde{\Sigma}^{ggg}_{(1)}(p,q,0,0) \nonumber \\
&& + \left[ \left[ \frac{7}{4} + \frac{14\pi^2}{27} 
- \frac{7}{9} \psi^\prime \left( \frac{1}{3} \right) \right] 
\frac{\muX^2}{\mu^2} \right. \nonumber \\
&& \left. ~~~+ \left[ \frac{5\pi^2}{27} - \frac{17}{12} - \frac{5}{18} 
\psi^\prime \left( \frac{1}{3} \right) + \frac{49}{16} \ln \left[ 
\frac{\muX^2}{\mu^2} \right] \right] \frac{\muX^4}{\mu^4} 
+ O \left( \frac{\muX^6}{\mu^6} \right) \right] C_A a \nonumber \\
&& +~ O(a^2) \nonumber \\
\tilde{\Sigma}^{ggg}_{(2)}(p,q,0,\muX^2) &=& 
\tilde{\Sigma}^{ggg}_{(2)}(p,q,0,0) \nonumber \\
&& + \left[ \left[ \frac{673}{96} - \frac{4\pi^2}{27} + \frac{2}{9} 
\psi^\prime \left( \frac{1}{3} \right) - \frac{3}{16} \ln \left[ 
\frac{\muX^2}{\mu^2} \right] \right] \frac{\muX^2}{\mu^2} \right. \nonumber \\
&& \left. ~~~+ \left[ \frac{209}{36} - \frac{26}{3} \ln \left[ 
\frac{\muX^2}{\mu^2} \right] \right] \frac{\muX^4}{\mu^4} 
+ O \left( \frac{\muX^6}{\mu^6} \right) \right] C_A a ~+~ O(a^2) \nonumber \\
\tilde{\Sigma}^{ggg}_{(3)}(p,q,0,\muX^2) &=& 
\tilde{\Sigma}^{ggg}_{(3)}(p,q,0,0) \nonumber \\
&& + \left[ \left[ \frac{596}{96} - \frac{4\pi^2}{9} + \frac{2}{3} 
\psi^\prime \left( \frac{1}{3} \right) - \frac{3}{16} \ln \left[ 
\frac{\muX^2}{\mu^2} \right] \right] \frac{\muX^2}{\mu^2} \right. \nonumber \\
&& \left. ~~~+ \left[ \frac{433}{72} - \frac{8\pi^2}{27} + \frac{4}{9} 
\psi^\prime \left( \frac{1}{3} \right) - \frac{211}{24} \ln \left[ 
\frac{\muX^2}{\mu^2} \right] \right] \frac{\muX^4}{\mu^4} 
+ O \left( \frac{\muX^6}{\mu^6} \right) \right] C_A a \nonumber \\
&& +~ O(a^2) ~. 
\end{eqnarray}
In contrast to the pure Gribov-Zwanziger case and the ${\cal Q}$ and 
${\cal R}$ extensions there is a dimension two correction in channel $1$ as 
well as the other two channels. This is a significant departure from the three 
Gribov scenarios and will form part of our test. For the other two vertices we 
have 
\begin{eqnarray}
\Sigma^{ccg}_{(1)}(p,q,0,\muX^2) &=& 
\Sigma^{ccg}_{(1)}(p,q,0,0) \nonumber \\
&& + \left[ \left[ \frac{1}{36} \psi^\prime \left( \frac{1}{3} \right) 
- \frac{11}{32} - \frac{\pi^2}{54} + \frac{9}{16} \ln \left[ 
\frac{\muX^2}{\mu^2} \right] \right] \frac{\muX^2}{\mu^2} \right. \nonumber \\
&& \left. ~~~+ \left[ \frac{83}{576} - \frac{\pi^2}{54} + \frac{1}{36}
\psi^\prime \left( \frac{1}{3} \right) + \frac{5}{24} \ln \left[ 
\frac{\muX^2}{\mu^2} \right] \right] \! \frac{\muX^4}{\mu^4} 
+ O \left( \frac{\muX^6}{\mu^6} \right) \right] C_A a \nonumber \\
&& +~ O(a^2) \nonumber \\
\Sigma^{ccg}_{(2)}(p,q,0,\muX^2) &=& 
\Sigma^{ccg}_{(2)}(p,q,0,0) \nonumber \\
&& + \left[ \left[ \frac{1}{8} - \frac{\pi^2}{27} + \frac{1}{18} \psi^\prime 
\left( \frac{1}{3} \right) - \frac{3}{8} \ln \left[ \frac{\muX^2}{\mu^2} 
\right] \right] \frac{\muX^2}{\mu^2} \right. \nonumber \\
&& \left. ~~~+ \left[ \frac{13}{576} + \frac{\pi^2}{54} - \frac{1}{36}
\psi^\prime \left( \frac{1}{3} \right) - \frac{1}{3} \ln \left[ 
\frac{\muX^2}{\mu^2} \right] \right] \frac{\muX^4}{\mu^4} \right. \nonumber \\
&& \left. +~ O \left( \frac{\muX^6}{\mu^6} \right) \right] C_A a ~+~ O(a^2) 
\end{eqnarray}
and
\begin{eqnarray}
\Sigma^{qqg}_{(1)}(p,q,0,\muX^2) &=& 
\Sigma^{qqg}_{(1)}(p,q,0,0) \nonumber \\
&& + \left[ \left[ \left[ \frac{13}{18} \psi^\prime \left( \frac{1}{3} \right) 
+ \frac{5}{16} - \frac{13\pi^2}{27} - \frac{3}{8} \ln \left[ 
\frac{\muX^2}{\mu^2} \right] \right] C_A \right. \right. \nonumber \\
&& \left. \left. ~~~~~+
\left[ \frac{7}{8} - \frac{5}{9} \psi^\prime \left( \frac{1}{3} \right) 
+ \frac{10\pi^2}{27} - \frac{3}{4} \ln \left[ \frac{\muX^2}{\mu^2} \right] 
\right] C_F \right] \frac{\muX^2}{\mu^2} \right. \nonumber \\
&& \left. ~~~+ \left[ \left[ \frac{661}{288} + \frac{8\pi^2}{27} - \frac{4}{9}
\psi^\prime \left( \frac{1}{3} \right) - \frac{49}{24} \ln \left[ 
\frac{\muX^2}{\mu^2} \right] \right] C_A \right. \right. \nonumber \\
&& \left. \left. ~~~~~~~+
\left[ \frac{4}{9} \psi^\prime \left( \frac{1}{3} \right) - \frac{67}{36}
- \frac{8\pi^2}{27} + \frac{4}{3} \ln \left[ \frac{\muX^2}{\mu^2} \right] 
\right] C_F \right] \frac{\muX^4}{\mu^4} 
+ O \left( \frac{\muX^6}{\mu^6} \right) \right] a \nonumber \\
&& +~ O(a^2) \nonumber \\
\Sigma^{qqg}_{(2)}(p,q,0,\muX^2) &=& \Sigma^{qqg}_{(5)}(p,q,0,\muX^2)
\nonumber \\
&=& \Sigma^{qqg}_{(2)}(p,q,0,0) \nonumber \\
&& + \left[ \left[ \left[ \frac{8}{9} \psi^\prime \left( \frac{1}{3} \right) 
+ \frac{7}{48} - \frac{16\pi^2}{27} + \frac{3}{8} \ln \left[ 
\frac{\muX^2}{\mu^2} \right] \right] C_A \right. \right. \nonumber \\
&& \left. \left. ~~~~~+
\left[ \frac{5}{3} - \frac{8}{9} \psi^\prime \left( \frac{1}{3} \right) 
+ \frac{16\pi^2}{27} 
\right] C_F \right] \frac{\muX^2}{\mu^2} \right. \nonumber \\
&& \left. ~~+ \left[ \left[ \frac{295}{72} + \frac{16\pi^2}{27} - \frac{8}{9}
\psi^\prime \left( \frac{1}{3} \right) - \frac{79}{24} \ln \left[ 
\frac{\muX^2}{\mu^2} \right] \right] C_A \right. \right. \nonumber \\
&& \left. \left. ~~~~~~~+
\left[ \frac{8}{9} \psi^\prime \left( \frac{1}{3} \right) - \frac{43}{18}
- \frac{16\pi^2}{27} + \frac{5}{3} \ln \left[ \frac{\muX^2}{\mu^2} \right] 
\right] C_F \right] \frac{\muX^4}{\mu^4} 
+ O \left( \frac{\muX^6}{\mu^6} \right) \right] a \nonumber \\
&& +~ O(a^2) \nonumber \\
\Sigma^{qqg}_{(3)}(p,q,0,\muX^2) &=& \Sigma^{qqg}_{(4)}(p,q,0,\muX^2)
\nonumber \\
&=& \Sigma^{qqg}_{(3)}(p,q,0,0) \nonumber \\
&& + \left[ \left[ \left[ \frac{2}{9} \psi^\prime \left( \frac{1}{3} \right) 
+ \frac{29}{48} - \frac{4\pi^2}{27} - \frac{3}{8} \ln \left[ 
\frac{\muX^2}{\mu^2} \right] \right] C_A \right. \right. \nonumber \\
&& \left. \left. ~~~~~+
\left[ \frac{19}{12} - \frac{2}{9} \psi^\prime \left( \frac{1}{3} \right) 
+ \frac{4\pi^2}{27} + \frac{3}{2} \ln \left[ \frac{\muX^2}{\mu^2} \right] 
\right] C_F \right] \frac{\muX^2}{\mu^2} \right. \nonumber \\
&& \left. ~~+ \left[ \left[ \frac{65}{144} - \frac{25}{24} \ln \left[ 
\frac{\muX^2}{\mu^2} \right] \right] C_A \right. \right. \nonumber \\
&& \left. \left. ~~~~~~~+ \left[ \frac{4}{3} - \ln \left[ \frac{\muX^2}{\mu^2} 
\right] \right] C_F \right] \frac{\muX^4}{\mu^4} 
+ O \left( \frac{\muX^6}{\mu^6} \right) \right] a \,+\, O(a^2) \nonumber \\
\Sigma^{qqg}_{(6)}(p,q,0,\muX^2) &=& 
\Sigma^{qqg}_{(6)}(p,q,0,0) \nonumber \\
&& + \left[ \left[ \left[ \frac{7}{9} \psi^\prime \left( \frac{1}{3} \right) 
- \frac{27}{8} - \frac{14\pi^2}{27} + \frac{9}{4} \ln \left[ 
\frac{\muX^2}{\mu^2} \right] \right] C_A \right. \right. \nonumber \\
&& \left. \left. ~~~~~+
\left[ \frac{7}{4} - \frac{2}{3} \psi^\prime \left( \frac{1}{3} \right) 
+ \frac{4\pi^2}{9} - \frac{3}{2} \ln \left[ \frac{\muX^2}{\mu^2} \right] 
\right] C_F \right] \frac{\muX^2}{\mu^2} \right. \nonumber \\
&& \left. ~~+ \left[ \left[ \frac{223}{144} + \frac{17}{12} \ln \left[ 
\frac{\muX^2}{\mu^2} \right] \right] C_A \right. \right. \nonumber \\
&& \left. \left. ~~~~~~~+
\left[ \frac{2}{3} \ln \left[ \frac{\muX^2}{\mu^2} \right] - \frac{31}{18}  
\right] C_F \right] \frac{\muX^4}{\mu^4} 
+ O \left( \frac{\muX^6}{\mu^6} \right) \right] a ~+~ O(a^2) ~.
\end{eqnarray}
Both these cases are completely parallel to earlier sections. 

Considering the situation where there is a Gribov mass as well as a gluon mass 
term we have
\begin{eqnarray}
\tilde{\Sigma}^{ggg}_{(1)}(p,q,\gamma^2,\muX^2) &=& 
\tilde{\Sigma}^{ggg}_{(1)}(p,q,0,0) \nonumber \\
&& + \left[ \left[ \frac{7}{4} + \frac{14\pi^2}{27} 
- \frac{7}{9} \psi^\prime \left( \frac{1}{3} \right) \right] 
\frac{C_A \muX^2}{\mu^2} \right. \nonumber \\
&& \left. ~~~+
\left[ \frac{57}{64} \muX^2 \sqrt{[\muX^4-4C_A\gamma^4]} 
\ln \left[ \frac{\muXp^2}{\muXm^2} \right] 
\right. \right. \nonumber \\
&& \left. \left. ~~~~~~~+ 
\left[ \frac{5\pi^2}{27} - \frac{17}{12} - \frac{5}{18} \psi^\prime 
\left( \frac{1}{3} \right) + \frac{49}{32} \ln 
\left[ \frac{C_A\gamma^4}{\mu^4} \right] \right] \muX^4 
\right. \right. \nonumber \\
&& \left. \left. ~~~~~~~+ 
\frac{41\muX^6}{64\sqrt{[\muX^4-4C_A\gamma^4]}}
\ln \left[ \frac{\muXp^2}{\muXm^2} \right] 
\right. \right. \nonumber \\
&& \left. \left. ~~~~~~~+ 
\left[ \frac{13}{6} + \frac{7\pi^2}{36} - \frac{7}{24} \psi^\prime 
\left( \frac{1}{3} \right) - \frac{1}{2} \ln \left[ \frac{C_A\gamma^4}{\mu^4} 
\right] \right] C_A\gamma^4 \right] \frac{C_A}{\mu^4} \right. \nonumber \\
&& \left. ~~~~~~~+ O \left( \frac{\muX^6}{\mu^6} \right) \right] a ~+~ O(a^2) 
\nonumber \\ 
\tilde{\Sigma}^{ggg}_{(2)}(p,q,\gamma^2,\muX^2) &=& 
\tilde{\Sigma}^{ggg}_{(2)}(p,q,0,0) \nonumber \\
&& + \left[ \left[ \left[ \frac{673}{96} - \frac{4\pi^2}{27} 
+ \frac{2}{9} \psi^\prime \left( \frac{1}{3} \right) 
- \frac{3}{32} \ln \left[ \frac{C_A\gamma^4}{\mu^4} \right] \right] \muX^2
\right. \right. \nonumber \\
&& \left. \left. ~~~~~~~- 
\frac{3\muX^4}{64\sqrt{[\muX^4-4C_A\gamma^4]}}
\ln \left[ \frac{\muXp^2}{\muXm^2} \right] 
\right. \right. \nonumber \\
&& \left. \left. ~~~~~~~- 
\frac{3}{64} \sqrt{[\muX^4-4C_A\gamma^4]}
\ln \left[ \frac{\muXp^2}{\muXm^2} \right] \right] 
\frac{C_A}{\mu^2} \right. \nonumber \\
&& \left. ~~~+
\left[ -~ \frac{817}{256} \muX^2 \sqrt{[\muX^4-4C_A\gamma^4]} 
\ln \left[ \frac{\muXp^2}{\muXm^2} \right] 
+ \left[ \frac{209}{36} - \frac{13}{3} \ln \left[ \frac{C_A\gamma^4}{\mu^4} 
\right] \right] \muX^4 
\right. \right. \nonumber \\
&& \left. \left. ~~~~~~~- 
\frac{877\muX^6}{768\sqrt{[\muX^4-4C_A\gamma^4]}}
\ln \left[ \frac{\muXp^2}{\muXm^2} \right] 
\right. \right. \nonumber \\
&& \left. \left. ~~~~~~~+ 
\left[ \frac{289}{144} 
+ \frac{787}{192} \ln \left[ \frac{C_A\gamma^4}{\mu^4} 
\right] \right] C_A\gamma^4 \right] \frac{C_A}{\mu^4} 
+ O \left( \frac{\muX^6}{\mu^6} \right) \right] a \nonumber \\
&& +~ O(a^2) \nonumber \\ 
\tilde{\Sigma}^{ggg}_{(3)}(p,q,\gamma^2,\muX^2) &=& 
\tilde{\Sigma}^{ggg}_{(3)}(p,q,0,0) \nonumber \\
&& + \left[ \left[ \left[ \frac{596}{96} - \frac{4\pi^2}{9} 
+ \frac{2}{3} \psi^\prime \left( \frac{1}{3} \right) 
- \frac{3}{32} \ln \left[ \frac{C_A\gamma^4}{\mu^4} \right] \right] \muX^2
\right. \right. \nonumber \\
&& \left. \left. ~~~~~~~- 
\frac{3\muX^4}{64\sqrt{[\muX^4-4C_A\gamma^4]}}
\ln \left[ \frac{\muXp^2}{\muXm^2} \right] 
\right. \right. \nonumber \\
&& \left. \left. ~~~~~~~- 
\frac{3}{64} \sqrt{[\muX^4-4C_A\gamma^4]}
\ln \left[ \frac{\muXp^2}{\muXm^2} \right] \right] 
\frac{C_A}{\mu^2} \right. \nonumber \\
&& \left. ~~~+
\left[ -~ \frac{829}{256} \muX^2 \sqrt{[\muX^4-4C_A\gamma^4]} 
\ln \left[ \frac{\muXp^2}{\muXm^2} \right] 
\right. \right. \nonumber \\
&& \left. \left. ~~~~~~~+ 
\left[ \frac{433}{72} - \frac{8\pi^2}{27} + \frac{4}{9} \psi^\prime 
\left( \frac{1}{3} \right) - \frac{211}{48} \ln \left[ 
\frac{C_A\gamma^4}{\mu^4} \right] \right] \muX^4 
\right. \right. \nonumber \\
&& \left. \left. ~~~~~~~- 
\frac{889\muX^6}{768\sqrt{[\muX^4-4C_A\gamma^4]}}
\ln \left[ \frac{\muXp^2}{\muXm^2} \right] 
\right. \right. \nonumber \\
&& \left. \left. ~~~~~~~+ 
\left[ \frac{25}{18} \psi^\prime \left( \frac{1}{3} \right) - \frac{2033}{576} 
- \frac{25\pi^2}{27} + \frac{799}{192} \ln \left[ \frac{C_A\gamma^4}{\mu^4} 
\right] \right] C_A\gamma^4 \right] \frac{C_A}{\mu^4} \right. \nonumber \\
&& \left.  
+~ O \left( \frac{\muX^6}{\mu^6} \right) \right] a ~+~ O(a^2) 
\end{eqnarray}
for the triple gluon vertex. The dimension two contribution in channel $1$ 
remains as the leading correction but interestingly it is independent of 
$\gamma$. This is not the case for the other two channels in that the dimension
two term derives from a combination of the two basic scales $\muX$ and 
$\gamma$. So it would appear that at one loop there is a clear way of
distinguishing between the masses in the problem. If a lattice computation 
found a dimension two contribution in channel $1$ then that would be a clear 
indication of an explicit gluon mass term. Though it would not determine 
whether or not there was also a Gribov mass present too. That could be deduced 
from the details of other channels and vertices. For the ghost-gluon and 
quark-gluon vertices we have 
\begin{eqnarray}
\Sigma^{ccg}_{(1)}(p,q,\gamma^2,\muX^2) &=& 
\Sigma^{ccg}_{(1)}(p,q,0,0) \nonumber \\
&& + \left[ \left[ \left[ \frac{1}{36} \psi^\prime \left( \frac{1}{3} \right) 
- \frac{11}{32} - \frac{\pi^2}{54} 
+ \frac{9}{32} \ln \left[ \frac{C_A\gamma^4}{\mu^4} \right] \right] \muX^2
\right. \right. \nonumber \\
&& \left. \left. ~~~~~~~+
\frac{9}{32} \sqrt{[\muX^4-4C_A\gamma^4]} \ln \left[ \frac{\muXp^2}{\muXm^2} 
\right] 
\right. \right. \nonumber \\
&& \left. \left. ~~~~~~~+
\frac{9C_A\gamma^4}{16\sqrt{[\muX^4-4C_A\gamma^4]}} \ln 
\left[ \frac{\muXp^2}{\muXm^2} \right] \right] \frac{C_A}{\mu^2} 
\right. \nonumber \\
&& \left. ~~~+
\left[ \frac{5}{48} \muX^2 \sqrt{[\muX^4-4C_A\gamma^4]} 
\ln \left[ \frac{\muXp^2}{\muXm^2} \right] 
\right. \right. \nonumber \\
&& \left. \left. ~~~~~~~+ 
\left[ \frac{83}{576} - \frac{\pi^2}{54}
+ \frac{1}{36} \psi^\prime \left( \frac{1}{3} \right) 
+ \frac{5}{48} \ln \left[ \frac{C_A\gamma^4}{\mu^4} 
\right] \right] \muX^4 
\right. \right. \nonumber \\
&& \left. \left. ~~~~~~~+
\frac{7C_A\muX^2\gamma^4}{24\sqrt{[\muX^4-4C_A\gamma^4]}}
\ln \left[ \frac{\muXp^2}{\muXm^2} \right] 
\right. \right. \nonumber \\
&& \left. \left. ~~~~~~~+ 
\left[ \frac{\pi^2}{216} - \frac{377}{2304}
- \frac{1}{144} \psi^\prime \left( \frac{1}{3} \right) 
+ \frac{1}{12} \ln \left[ \frac{C_A\gamma^4}{\mu^4} 
\right] \right] C_A\gamma^4 \right] \frac{C_A}{\mu^4} \right. \nonumber \\
&& \left. +~ O \left( \frac{\muX^6}{\mu^6} \right) \right] a ~+~ O(a^2) 
\nonumber \\ 
\Sigma^{ccg}_{(2)}(p,q,\gamma^2,\muX^2) &=& 
\Sigma^{ccg}_{(2)}(p,q,0,0) \nonumber \\
&& + \left[ \left[ \left[ \frac{1}{18} \psi^\prime \left( \frac{1}{3} \right) 
+ \frac{1}{8} - \frac{\pi^2}{27} 
- \frac{3}{16} \ln \left[ \frac{C_A\gamma^4}{\mu^4} \right] \right] \muX^2
\right. \right. \nonumber \\
&& \left. \left. ~~~~~~~- 
\frac{3}{16} \sqrt{[\muX^4-4C_A\gamma^4]} \ln \left[ \frac{\muXp^2}{\muXm^2} 
\right] 
\right. \right. \nonumber \\
&& \left. \left. ~~~~~~~-
\frac{3C_A\gamma^4}{8\sqrt{[\muX^4-4C_A\gamma^4]}} \ln 
\left[ \frac{\muXp^2}{\muXm^2} \right] \right] \frac{C_A}{\mu^2} 
\right. \nonumber \\
&& \left. ~~~+
\left[ -~ \frac{1}{6} \muX^2 \sqrt{[\muX^4-4C_A\gamma^4]} 
\ln \left[ \frac{\muXp^2}{\muXm^2} \right] 
\right. \right. \nonumber \\
&& \left. \left. ~~~~~~~+ 
\left[ \frac{13}{576} + \frac{\pi^2}{54}
- \frac{1}{36} \psi^\prime \left( \frac{1}{3} \right) 
- \frac{1}{6} \ln \left[ \frac{C_A\gamma^4}{\mu^4} 
\right] \right] \muX^4 
\right. \right. \nonumber \\
&& \left. \left. ~~~~~~~- 
\frac{17C_A\muX^2\gamma^4}{48\sqrt{[\muX^4-4C_A\gamma^4]}}
\ln \left[ \frac{\muXp^2}{\muXm^2} \right] 
\right. \right. \nonumber \\
&& \left. \left. ~~~~~~~+ 
\left[ \frac{1}{144} \psi^\prime \left( \frac{1}{3} \right) 
- \frac{\pi^2}{216} - \frac{7}{2304}
- \frac{1}{48} \ln \left[ \frac{C_A\gamma^4}{\mu^4} 
\right] \right] C_A\gamma^4 \right] \frac{C_A}{\mu^4} \right. \nonumber \\
&& \left. +~ O \left( \frac{\muX^6}{\mu^6} \right) \right] a ~+~ O(a^2) 
\end{eqnarray}
and
\begin{eqnarray}
\Sigma^{qqg}_{(1)}(p,q,\gamma^2,\muX^2) &=& 
\Sigma^{qqg}_{(1)}(p,q,0,0) \nonumber \\
&& + \left[ \left[ \left[ 
-~ \frac{3C_A\gamma^4}{4\sqrt{[\muX^4-4C_A\gamma^4]}} 
\ln \left[ \frac{\muXp^2}{\muXm^2} \right] 
\right. \right. \right. \nonumber \\
&& \left. \left. \left. ~~~~~~~
+ \left[ \frac{7}{8} + \frac{10\pi^2}{27} - \frac{5}{9} \psi^\prime 
\left( \frac{1}{3} \right) - \frac{3}{8} \ln \left[ \frac{C_A\gamma^4}{\mu^4} 
\right] \right] \muX^2
\right. \right. \right. \nonumber \\
&& \left. \left. \left. ~~~~~~~
- \frac{3}{8} \sqrt{[\muX^4-4C_A\gamma^4]}
\ln \left[ \frac{\muXp^2}{\muXm^2} \right] \right] C_F
\right. \right. \nonumber \\
&& \left. \left. ~~~~~~
+ \left[ 
\left[ \frac{5}{16} - \frac{13\pi^2}{27} + \frac{13}{18} \psi^\prime 
\left( \frac{1}{3} \right) - \frac{3}{16} \ln \left[ \frac{C_A\gamma^4}{\mu^4} 
\right] \right] \muX^2
\right. \right. \right. \nonumber \\
&& \left. \left. \left. ~~~~~~~~~~
- \frac{3}{16} \sqrt{[\muX^4-4C_A\gamma^4]}
\ln \left[ \frac{\muXp^2}{\muXm^2} \right] 
\right. \right. \right. \nonumber \\
&& \left. \left. \left. ~~~~~~~~~~
- \frac{3C_A\gamma^4}{8\sqrt{[\muX^4-4C_A\gamma^4]}} 
\ln \left[ \frac{\muXp^2}{\muXm^2} \right] \right] C_A 
\right] \frac{1}{\mu^2} 
\right. \nonumber \\
&& \left. ~~~
+ \left[ \left[ 
\frac{2C_A\muX^2\gamma^4}{3\sqrt{[\muX^4-4C_A\gamma^4]}} 
\ln \left[ \frac{\muXp^2}{\muXm^2} \right] 
\right. \right. \right. \nonumber \\
&& \left. \left. \left. ~~~~~~~
+ \left[ \frac{67}{36} + \frac{8\pi^2}{27} - \frac{4}{9} \psi^\prime 
\left( \frac{1}{3} \right) - \frac{2}{3} \ln \left[ \frac{C_A\gamma^4}{\mu^4} 
\right] \right] C_A \gamma^4
\right. \right. \right. \nonumber \\
&& \left. \left. \left. ~~~~~~~
+ \frac{2}{3} \muX^2 \sqrt{[\muX^4-4C_A\gamma^4]}
\ln \left[ \frac{\muXp^2}{\muXm^2} \right]
\right. \right. \right. \nonumber \\
&& \left. \left. \left. ~~~~~~~
+ \left[ \frac{4}{9} \psi^\prime \left( \frac{1}{3} \right) - \frac{67}{36} 
- \frac{8\pi^2}{27} + \frac{2}{3} \ln \left[ \frac{C_A\gamma^4}{\mu^4} 
\right] \right] \muX^4 \right] C_F
\right. \right. \nonumber \\
&& \left. \left. ~~~
+ \left[ 
-~ \frac{49}{48} \muX^2 \sqrt{[\muX^4-4C_A\gamma^4]}
\ln \left[ \frac{\muXp^2}{\muXm^2} \right] 
\right. \right. \right. \nonumber \\
&& \left. \left. \left. ~~~~~~~
+ \left[ \frac{661}{288} + \frac{8\pi^2}{27} - \frac{4}{9} \psi^\prime 
\left( \frac{1}{3} \right) - \frac{49}{48} \ln \left[ \frac{C_A\gamma^4}{\mu^4}
\right] \right] \muX^4
\right. \right. \right. \nonumber \\
&& \left. \left. \left. ~~~~~~~
- \frac{67C_A\muX^2\gamma^4}{48\sqrt{[\muX^4-4C_A\gamma^4]}} 
\ln \left[ \frac{\muXp^2}{\muXm^2} \right] 
\right. \right. \right. \nonumber \\
&& \left. \left. \left. ~~~~~~~
+ \left[ \frac{11}{18} \psi^\prime \left( \frac{1}{3} \right) 
- \frac{2941}{1152} - \frac{11\pi^2}{27} 
+ \frac{31}{48} \ln \left[ \frac{C_A\gamma^4}{\mu^4} 
\right] \right] C_A \gamma^4 \right] C_A \right] \frac{1}{\mu^4}
\right. \nonumber \\
&& \left. ~+ O \left( \frac{\muX^6}{\mu^6} \right) \right] a ~+~ O(a^2) 
\nonumber \\ 
\Sigma^{qqg}_{(2)}(p,q,\gamma^2,\muX^2) &=& 
\Sigma^{qqg}_{(5)}(p,q,\gamma^2,\muX^2) \nonumber \\
&=& \Sigma^{qqg}_{(2)}(p,q,0,0) \nonumber \\
&& + \left[ \left[ \left[ 
\left[ \frac{5}{3} + \frac{16\pi^2}{27} - \frac{8}{9} \psi^\prime 
\left( \frac{1}{3} \right) 
\right] \muX^2
\right] C_F
\right. \right. \nonumber \\
&& \left. \left. ~~~~~~
+ \left[ 
\left[ \frac{7}{48} - \frac{16\pi^2}{27} + \frac{8}{9} \psi^\prime 
\left( \frac{1}{3} \right) + \frac{3}{16} \ln \left[ \frac{C_A\gamma^4}{\mu^4} 
\right] \right] \muX^2
\right. \right. \right. \nonumber \\
&& \left. \left. \left. ~~~~~~~~~~
+ \frac{3}{16} \sqrt{[\muX^4-4C_A\gamma^4]}
\ln \left[ \frac{\muXp^2}{\muXm^2} \right] 
\right. \right. \right. \nonumber \\
&& \left. \left. \left. ~~~~~~~~~~
+ \frac{3C_A\gamma^4}{8\sqrt{[\muX^4-4C_A\gamma^4]}} 
\ln \left[ \frac{\muXp^2}{\muXm^2} \right] \right] C_A 
\right] \frac{1}{\mu^2} 
\right. \nonumber \\
&& \left. ~~~
+ \left[ \left[ 
\frac{5C_A\muX^2\gamma^4}{6\sqrt{[\muX^4-4C_A\gamma^4]}} 
\ln \left[ \frac{\muXp^2}{\muXm^2} \right] 
\right. \right. \right. \nonumber \\
&& \left. \left. \left. ~~~~~~~
+ \left[ \frac{43}{18} + \frac{16\pi^2}{27} - \frac{8}{9} \psi^\prime 
\left( \frac{1}{3} \right) - \frac{5}{6} \ln \left[ \frac{C_A\gamma^4}{\mu^4} 
\right] \right] C_A \gamma^4
\right. \right. \right. \nonumber \\
&& \left. \left. \left. ~~~~~~~
+ \frac{5}{6} \muX^2 \sqrt{[\muX^4-4C_A\gamma^4]}
\ln \left[ \frac{\muXp^2}{\muXm^2} \right]
\right. \right. \right. \nonumber \\
&& \left. \left. \left. ~~~~~~
+ \left[ \frac{8}{9} \psi^\prime \left( \frac{1}{3} \right) - \frac{43}{18} 
- \frac{16\pi^2}{27} + \frac{5}{6} \ln \left[ \frac{C_A\gamma^4}{\mu^4} 
\right] \right] \muX^4 \right] C_F
\right. \right. \nonumber \\
&& \left. \left. ~~~
+ \left[ 
-~ \frac{79}{48} \muX^2 \sqrt{[\muX^4-4C_A\gamma^4]}
\ln \left[ \frac{\muXp^2}{\muXm^2} \right] 
\right. \right. \right. \nonumber \\
&& \left. \left. \left. ~~~~~~~
+ \left[ \frac{295}{72} + \frac{16\pi^2}{27} - \frac{8}{9} \psi^\prime 
\left( \frac{1}{3} \right) - \frac{79}{48} \ln \left[ \frac{C_A\gamma^4}{\mu^4}
\right] \right] \muX^4
\right. \right. \right. \nonumber \\
&& \left. \left. \left. ~~~~~~~
- \frac{307C_A\muX^2\gamma^4}{192\sqrt{[\muX^4-4C_A\gamma^4]}} 
\ln \left[ \frac{\muXp^2}{\muXm^2} \right] 
\right. \right. \right. \nonumber \\
&& \left. \left. \left. ~~~~~~~
+ \left[ \frac{14}{9} \psi^\prime \left( \frac{1}{3} \right) 
- \frac{1351}{288} - \frac{28\pi^2}{27} 
+ \frac{325}{192} \ln \left[ \frac{C_A\gamma^4}{\mu^4} 
\right] \right] C_A \gamma^4 \right] C_A \right] \frac{1}{\mu^4}
\right. \nonumber \\
&& \left. ~+ O \left( \frac{\muX^6}{\mu^6} \right) \right] a ~+~ O(a^2) 
\nonumber \\ 
\Sigma^{qqg}_{(3)}(p,q,\gamma^2,\muX^2) &=& 
\Sigma^{qqg}_{(4)}(p,q,\gamma^2,\muX^2) \nonumber \\
&=& \Sigma^{qqg}_{(3)}(p,q,0,0) \nonumber \\
&& + \left[ \left[ \left[ 
\frac{3C_A\gamma^4}{2\sqrt{[\muX^4-4C_A\gamma^4]}} 
\ln \left[ \frac{\muXp^2}{\muXm^2} \right] 
\right. \right. \right. \nonumber \\
&& \left. \left. \left. ~~~~~~~
+ \left[ \frac{19}{12} + \frac{4\pi^2}{27} - \frac{2}{9} \psi^\prime 
\left( \frac{1}{3} \right) + \frac{3}{4} \ln \left[ \frac{C_A\gamma^4}{\mu^4} 
\right] \right] \muX^2
\right. \right. \right. \nonumber \\
&& \left. \left. \left. ~~~~~~~
+ \frac{3}{4} \sqrt{[\muX^4-4C_A\gamma^4]}
\ln \left[ \frac{\muXp^2}{\muXm^2} \right] \right] C_F
\right. \right. \nonumber \\
&& \left. \left. ~~~~~~
+ \left[ 
\left[ \frac{29}{48} - \frac{4\pi^2}{27} + \frac{2}{9} \psi^\prime 
\left( \frac{1}{3} \right) - \frac{3}{16} \ln \left[ \frac{C_A\gamma^4}{\mu^4} 
\right] \right] \muX^2
\right. \right. \right. \nonumber \\
&& \left. \left. \left. ~~~~~~~~~~
- \frac{3}{16} \sqrt{[\muX^4-4C_A\gamma^4]}
\ln \left[ \frac{\muXp^2}{\muXm^2} \right] 
\right. \right. \right. \nonumber \\
&& \left. \left. \left. ~~~~~~~~~~
- \frac{3C_A\gamma^4}{8\sqrt{[\muX^4-4C_A\gamma^4]}} 
\ln \left[ \frac{\muXp^2}{\muXm^2} \right] \right] C_A 
\right] \frac{1}{\mu^2} 
\right. \nonumber \\
&& \left. ~~~
+ \left[ \left[ 
-~ \frac{C_A\muX^2\gamma^4}{2\sqrt{[\muX^4-4C_A\gamma^4]}} 
\ln \left[ \frac{\muXp^2}{\muXm^2} \right] 
+ \left[ \frac{1}{2} \ln \left[ \frac{C_A\gamma^4}{\mu^4} \right]
- \frac{4}{3} \right] C_A \gamma^4
\right. \right. \right. \nonumber \\
&& \left. \left. \left. ~~~~~~~
- \frac{1}{2} \muX^2 \sqrt{[\muX^4-4C_A\gamma^4]}
\ln \left[ \frac{\muXp^2}{\muXm^2} \right]
+ \left[ \frac{4}{3} - \frac{1}{2} \ln \left[ \frac{C_A\gamma^4}{\mu^4} \right]
\right] \muX^4 \right] C_F
\right. \right. \nonumber \\
&& \left. \left. ~~~
+ \left[ 
-~ \frac{25}{48} \muX^2 \sqrt{[\muX^4-4C_A\gamma^4]}
\ln \left[ \frac{\muXp^2}{\muXm^2} \right] 
\right. \right. \right. \nonumber \\
&& \left. \left. \left. ~~~~~~~
+ \left[ \frac{65}{144} 
- \frac{25}{48} \ln \left[ \frac{C_A\gamma^4}{\mu^4}
\right] \right] \muX^4
- \frac{91C_A\muX^2\gamma^4}{192\sqrt{[\muX^4-4C_A\gamma^4]}} 
\ln \left[ \frac{\muXp^2}{\muXm^2} \right] 
\right. \right. \right. \nonumber \\
&& \left. \left. \left. ~~~~~~~
+ \left[ \frac{2}{3} \psi^\prime \left( \frac{1}{3} \right) 
- \frac{917}{576} - \frac{4\pi^2}{9} 
+ \frac{109}{192} \ln \left[ \frac{C_A\gamma^4}{\mu^4} 
\right] \right] C_A \gamma^4 \right] C_A \right] \frac{1}{\mu^4}
\right. \nonumber \\
&& \left. ~+ O \left( \frac{\muX^6}{\mu^6} \right) \right] a ~+~ O(a^2) 
\nonumber \\ 
\Sigma^{qqg}_{(6)}(p,q,\gamma^2,\muX^2) &=& 
\Sigma^{qqg}_{(6)}(p,q,0,0) \nonumber \\
&& + \left[ \left[ \left[ 
-~ \frac{3C_A\gamma^4}{2\sqrt{[\muX^4-4C_A\gamma^4]}} 
\ln \left[ \frac{\muXp^2}{\muXm^2} \right] 
\right. \right. \right. \nonumber \\
&& \left. \left. \left. ~~~~~~~
+ \left[ \frac{7}{4} + \frac{4\pi^2}{9} - \frac{2}{3} \psi^\prime 
\left( \frac{1}{3} \right) - \frac{3}{4} \ln \left[ \frac{C_A\gamma^4}{\mu^4} 
\right] \right] \muX^2
\right. \right. \right. \nonumber \\
&& \left. \left. \left. ~~~~~~~
-~ \frac{3}{4} \sqrt{[\muX^4-4C_A\gamma^4]}
\ln \left[ \frac{\muXp^2}{\muXm^2} \right] \right] C_F
\right. \right. \nonumber \\
&& \left. \left. ~~~~~~
+ \left[ 
\left[ \frac{7}{9} \psi^\prime \left( \frac{1}{3} \right) 
- \frac{27}{8} - \frac{14\pi^2}{27} 
+ \frac{9}{8} \ln \left[ \frac{C_A\gamma^4}{\mu^4} 
\right] \right] \muX^2
\right. \right. \right. \nonumber \\
&& \left. \left. \left. ~~~~~~~~~~
+ \frac{9}{8} \sqrt{[\muX^4-4C_A\gamma^4]}
\ln \left[ \frac{\muXp^2}{\muXm^2} \right] 
\right. \right. \right. \nonumber \\
&& \left. \left. \left. ~~~~~~~~~~
+ \frac{9C_A\gamma^4}{4\sqrt{[\muX^4-4C_A\gamma^4]}} 
\ln \left[ \frac{\muXp^2}{\muXm^2} \right] \right] C_A 
\right] \frac{1}{\mu^2} 
\right. \nonumber \\
&& \left. ~~~
+ \left[ \left[ 
\frac{C_A\muX^2\gamma^4}{3\sqrt{[\muX^4-4C_A\gamma^4]}} 
\ln \left[ \frac{\muXp^2}{\muXm^2} \right] 
+ \left[ \frac{31}{18} 
- \frac{1}{3} \ln \left[ \frac{C_A\gamma^4}{\mu^4} 
\right] \right] C_A \gamma^4
\right. \right. \right. \nonumber \\
&& \left. \left. \left. ~~~~~~~
+ \frac{1}{3} \muX^2 \sqrt{[\muX^4-4C_A\gamma^4]}
\ln \left[ \frac{\muXp^2}{\muXm^2} \right]
+ \left[ \frac{1}{3} \ln \left[ \frac{C_A\gamma^4}{\mu^4} \right]
- \frac{31}{18} \right] \muX^4 \right] C_F
\right. \right. \nonumber \\
&& \left. \left. ~~~
+ \left[ 
\frac{17}{24} \muX^2 \sqrt{[\muX^4-4C_A\gamma^4]}
\ln \left[ \frac{\muXp^2}{\muXm^2} \right] 
+ \left[ \frac{223}{144} 
+ \frac{17}{24} \ln \left[ \frac{C_A\gamma^4}{\mu^4}
\right] \right] \muX^4
\right. \right. \right. \nonumber \\
&& \left. \left. \left. ~~~~~~~
+ \frac{35C_A\muX^2\gamma^4}{24\sqrt{[\muX^4-4C_A\gamma^4]}} 
\ln \left[ \frac{\muXp^2}{\muXm^2} \right] 
\right. \right. \right. \nonumber \\
&& \left. \left. \left. ~~~~~~~
+ \left[ \frac{1}{24} \ln \left[ \frac{C_A\gamma^4}{\mu^4} \right] 
- \frac{991}{576} \right] C_A \gamma^4 \right] C_A \right] \frac{1}{\mu^4}
+ O \left( \frac{\muX^6}{\mu^6} \right) \right] a \nonumber \\
&& +~ O(a^2) ~. 
\end{eqnarray}
These are considerably more complicated than their ${\cal Q}$ and ${\cal R}$
counterparts. However, the key point in this respect both here and in previous 
results sections is that it is the actual mass scale, no matter how complicated
it is in each underlying theory, which is ultimately what the lattice will 
observe. We note that as a check for each of the three vertices the pure gluon 
mass expressions are obtained in the $\gamma^2$~$\rightarrow$~$0$ limit in each
case.

\sect{Discussion.}

We conclude by recalling that we have analysed the one loop corrections to the
$3$-point vertices of QCD at the symmetric subtraction point in the Landau
gauge using the Gribov-Zwanziger Lagrangian and its extensions which 
incorporate the Gribov problem. The main motivation was to examine the power 
corrections to the amplitudes in order to see whether the leading correction 
was dimension two or four. While \cite{1} focused on an effective coupling 
constant derived from $2$ and $3$-point functions and examined the deviations 
from expected behaviour, we have concentrated on just the vertices themselves. 
This is because the effective coupling constant definition involved the 
behaviour of the propagator form factors which have been shown in the 
Gribov-Zwanziger case to have dimension two corrections, \cite{51}. These would
therefore dominate in an effective coupling and the leading vertex correction 
behaviour would not be distinguishable. This is important since we have shown 
that in certain amplitudes the leading correction is dimension {\em four} and 
not dimension two. Indeed for any asymmetric momentum configuration for the 
three vertex functions the leading correction is always dimension two. The 
particular cases where dimension four is leading is in the channel of the 
triple gluon vertex which corresponds to the Feynman rule of the vertex itself.
This is the case not only for the pure Gribov-Zwanziger Lagrangian but also for
its extension to either what has been termed the ${\cal Q}$ and ${\cal R}$ 
solutions. For a simple gluon mass in the absence of the Gribov mass the triple
gluon vertex has a dimension two correction. Since we are at a symmetric 
subtraction point where the common scale of the external legs is not small then
examining the deviation from expected perturbative behaviour could provide an 
important test of the underlying mechanism. For instance, if in measuring the 
triple gluon vertex channel $1$ a power behaviour deviation of dimension two is
found then that would rule out a pure Gribov-Zwanziger or ${\cal Q}$ and 
${\cal R}$ explanation. It would not necessarily imply that a pure gluon mass 
is the underlying reason. This is because there are more complicated extensions
of (\ref{laggz}), \cite{30}, not considered here  where those propagators could 
mimic the dimension two correction. We have not introduced these here as we 
believe of the full set one of ${\cal Q}$ or ${\cal R}$ is naturally favourable
as discussed in \cite{30}. By contrast if a triple gluon vertex measurement 
indicated that the leading correction was clearly dimension four then that 
would suggest that the Gribov mass is playing a role. To decipher whether it is
the pure Gribov-Zwanziger case would require examination of the strength of the
relative corrections and also data on the other vertices. Though this clearly 
is at a level of fine detail. 

It is worth commenting on the current status of lattice measurements of the 
three vertices at the symmetric point. First, we note that there is only a 
small amount of data for this point compared with the asymmetric point. For the 
triple gluon vertex the original indication of a dimension two correction in 
the effective coupling constant was carried out in \cite{1}. However, in 
keeping with other analyses of this vertex and the other two it transpires that
signals suffer from more noise than the corresponding asymmetric vertex. This 
is despite the fact that the latter configuration requires a zero momentum 
limit. Therefore, at this stage one can not yet make any meaningful contact 
with data on this vertex function in order to see deviations from expected 
perturbative behaviour aside from the original observations of \cite{1}. 
Moreover, we note that in \cite{1} the vertex function for the triple gluon 
vertex was decomposed into only {\em two} independent tensors and not three as 
we have done here. Appendix A provides more details on this point. So even if a
comparison with data could be made with our power corrections it is not clear 
whether this would be meaningful given that the bases are different. Other 
studies of this vertex include \cite{52,53}. In the former the four dimensional
data, while noisy, show a general decrease towards zero momenta which is 
reinforced in the latter article. Thus again in these cases a direct comparison
with power corrections is not currently viable. For the ghost-gluon vertex the 
data of \cite{53} does not suffer from as much noise as the triple gluon vertex
case. There the main observation is that the ghost-gluon vertex is effectively 
equivalent to the tree value with a small maximum about $1$GeV, \cite{53}. For 
the quark-gluon vertex there is the additional issue of quenched versus 
dynamical data. The analysis of \cite{54} was in the quenched approximation and
at the symmetric point the data indicate a smoothly rising vertex function. 
Though the data is noisier than the asymmetric results presented there too and 
not sufficient in order to perform a comparison with power corrections. Despite
this one hope is that with the advances in lattice technology in recent years 
the focus could return to all of the $3$-point functions now that there seems 
to be a consensus on the zero momentum behaviour of the gluon and ghost 
$2$-point functions.

Finally, we should qualify our overall remarks by noting that we have performed
the analysis at one loop. One could regard this as a next to high energy 
expansion. However, there is no a priori reason why the leading dimension four 
correction of channel $1$ of the triple gluon vertex should persist beyond one 
loop. The leading two loop correction to this channel could be dimension two. 
So a more careful test could be that if the leading correction is dimension two
but relatively weak compared to the other leading one loop dimension two 
corrections in other channels then that could be evidence for a non-pure gluon 
mass explanation. To carry out a two loop extension to the expansion is in 
principle possible but is beyond the scope of the present paper. The technical 
calculational tools are clearly available. Though one would have a substantial 
number of {\sc Reduze} databases to build for the two main topologies that 
occur at two loops, \cite{38}, to cover all the non-zero mass distribution 
possibilities. While a one loop observation is by no means a proof it is 
intriguing that of the three $3$-point vertices of QCD it is actually the fully
symmetric one in terms of fields when examined in this power expansion 
specifically at the symmetric subtraction point which should have dimension 
four as the leading correction.

\vspace{1cm}
\noindent
{\bf Acknowledgement.} The author thanks Dr. P.A. Boyle for useful discussions.

\appendix

\sect{Tensor basis.}

In order to assist with the interpretation of the results in this appendix we
record the explicit forms of the tensors in the bases for each of the vertices.
For the triple gluon vertex the tensors of the original of \cite{38} are
\begin{eqnarray}
{\cal P}^{\mbox{\footnotesize{ggg}}}_{(1) \mu \nu \sigma }(p,q) &=&
\eta_{\mu \nu} p_\sigma ~~,~~
{\cal P}^{\mbox{\footnotesize{ggg}}}_{(2) \mu \nu \sigma }(p,q) ~=~
\eta_{\nu \sigma} p_\mu ~~,~~
{\cal P}^{\mbox{\footnotesize{ggg}}}_{(3) \mu \nu \sigma }(p,q) ~=~
\eta_{\sigma \mu} p_\nu \nonumber \\
{\cal P}^{\mbox{\footnotesize{ggg}}}_{(4) \mu \nu \sigma }(p,q) &=&
\eta_{\mu \nu} q_\sigma ~~,~~
{\cal P}^{\mbox{\footnotesize{ggg}}}_{(5) \mu \nu \sigma }(p,q) ~=~
\eta_{\nu \sigma} q_\mu ~~,~~
{\cal P}^{\mbox{\footnotesize{ggg}}}_{(6) \mu \nu \sigma }(p,q) ~=~
\eta_{\sigma \mu} q_\nu \nonumber \\
{\cal P}^{\mbox{\footnotesize{ggg}}}_{(7) \mu \nu \sigma }(p,q) &=&
\frac{1}{\mu^2} p_\mu p_\nu p_\sigma ~~,~~
{\cal P}^{\mbox{\footnotesize{ggg}}}_{(8) \mu \nu \sigma }(p,q) ~=~
\frac{1}{\mu^2} p_\mu p_\nu q_\sigma ~~,~~
{\cal P}^{\mbox{\footnotesize{ggg}}}_{(9) \mu \nu \sigma }(p,q) ~=~
\frac{1}{\mu^2} p_\mu q_\nu p_\sigma \nonumber \\
{\cal P}^{\mbox{\footnotesize{ggg}}}_{(10) \mu \nu \sigma }(p,q) &=&
\frac{1}{\mu^2} q_\mu p_\nu p_\sigma ~~,~~
{\cal P}^{\mbox{\footnotesize{ggg}}}_{(11) \mu \nu \sigma }(p,q) ~=~
\frac{1}{\mu^2} p_\mu q_\nu q_\sigma ~~,~~
{\cal P}^{\mbox{\footnotesize{ggg}}}_{(12) \mu \nu \sigma }(p,q) ~=~
\frac{1}{\mu^2} q_\mu p_\nu q_\sigma \nonumber \\
{\cal P}^{\mbox{\footnotesize{ggg}}}_{(13) \mu \nu \sigma }(p,q) &=&
\frac{1}{\mu^2} q_\mu q_\nu p_\sigma ~~,~~
{\cal P}^{\mbox{\footnotesize{ggg}}}_{(14) \mu \nu \sigma }(p,q) ~=~
\frac{1}{\mu^2} q_\mu q_\nu q_\sigma 
\end{eqnarray}
where the first six correspond to the terms of the original vertex in the 
Lagrangian. However, from explicit calculations, \cite{38}, it transpires that
the basis can be compactified into {\em three} independent combinations which 
we define to be
\begin{eqnarray}
\tilde{\cal P}^{ggg}_{(1) \mu \nu \sigma }(p,q) &=&
\eta_{\mu \nu} p_\sigma
- \eta_{\mu \nu} q_\sigma
- 2 \eta_{\mu \sigma} p_\nu
- \eta_{\sigma \mu} q_\nu
+ \eta_{\nu \sigma} p_\mu
+ 2 \eta_{\nu \sigma} q_\mu \nonumber \\
\tilde{\cal P}^{ggg}_{(2) \mu \nu \sigma }(p,q) &=&
\left[ 2 p_\mu p_\nu p_\sigma
+ p_\mu q_\nu p_\sigma
- p_\mu q_\nu q_\sigma 
+ 2 q_\mu p_\nu p_\sigma
- 2 q_\mu p_\nu q_\sigma
- 2 q_\mu q_\nu q_\sigma \right] \frac{1}{2\mu^2} \nonumber \\
\tilde{\cal P}^{ggg}_{(3) \mu \nu \sigma }(p,q) &=&
\left[
p_\mu p_\nu q_\sigma
- q_\mu p_\nu p_\sigma
+ q_\mu p_\nu q_\sigma
- q_\mu q_\nu p_\sigma \right] \frac{1}{\mu^2} ~. 
\label{projaaaalt}
\end{eqnarray}
The first again is the triple gluon vertex when one sets 
$r$~$=$~$-$~$p$~$-$~$q$. In the analysis of \cite{1} only two tensors are 
defined. One corresponds to the first and the second is a linear combination of
the other two. More specifically the second tensor of \cite{1} is proportional
to
\begin{equation}
\tilde{\cal P}^{ggg}_{(2) \mu \nu \sigma }(p,q) ~-~ 
\tilde{\cal P}^{ggg}_{(3) \mu \nu \sigma }(p,q) ~=~ -~ 
\frac{1}{2\mu^2} ( q - r )_\mu ( r - p )_\nu ( p - q )_\sigma ~.
\end{equation}  
Though the explicit two loop calculations of \cite{38} indicate that this 
choice of basis is too limited. In order to project out the amplitudes the 
projection matrix for (\ref{projaaaalt}) is
\begin{equation}
\tilde{\cal M}^{\mbox{\footnotesize{ggg}}}_{pq} ~=~ -~ \frac{1}{27(d-2)} \left(
\begin{array}{ccc}
3 & 0 & - 6 \\
0 & 16 (d-2) & 8 (d-2) \\
- 6 & 8 (d-2) & 4 (4d-5) \\
\end{array}
\right) 
\end{equation}
where the subscript denotes the symmetric point. The corresponding 
$14$~$\times$~$14$ matrix for 
${\cal P}^{\mbox{\footnotesize{ggg}}}_{(i) \mu \nu \sigma }(p,q)$ as well as
those for the quark and ghost vertices used here were given in \cite{38} but,
for completeness here, the tensors of the latter two respective bases are  
\begin{equation}
{\cal P}^{\mbox{\footnotesize{ccg}}}_{(1) \sigma }(p,q) ~=~ p_\sigma ~~~,~~~
{\cal P}^{\mbox{\footnotesize{ccg}}}_{(2) \sigma }(p,q) ~=~ q_\sigma 
\end{equation}
and
\begin{eqnarray}
{\cal P}^{\mbox{\footnotesize{qqg}}}_{(1) \sigma }(p,q) &=&
\gamma_\sigma ~~~,~~~
{\cal P}^{\mbox{\footnotesize{qqg}}}_{(2) \sigma }(p,q) ~=~
\frac{{p}_\sigma \pslash}{\mu^2} ~~~,~~~
{\cal P}^{\mbox{\footnotesize{qqg}}}_{(3) \sigma }(p,q) ~=~
\frac{{p}_\sigma \qslash}{\mu^2} ~, \nonumber \\
{\cal P}^{\mbox{\footnotesize{qqg}}}_{(4) \sigma }(p,q) &=&
\frac{{q}_\sigma \pslash}{\mu^2} ~~~,~~~
{\cal P}^{\mbox{\footnotesize{qqg}}}_{(5) \sigma }(p,q) ~=~
\frac{{q}_\sigma \qslash}{\mu^2} ~~~,~~~
{\cal P}^{\mbox{\footnotesize{qqg}}}_{(6) \sigma }(p,q) ~=~
\frac{1}{\mu^2} \Gamma_{(3) \, \sigma p q} ~.
\end{eqnarray}
We use the convention that when an external momentum is contracted with a
Lorentz index then the dummy index is replaced by that momentum.

\sect{Leading order amplitudes.}

In this appendix for completeness we record the explicit values for the various
one loop amplitudes of each vertex at the symmetric point in the absence of the
Gribov mass. They were computed originally in \cite{55}. We have
\begin{eqnarray}
\tilde{\Sigma}^{ggg}_{(1)}(p,q,0,0) &=& -~ 1
+ \left[ \left[ \frac{16}{27} \psi^\prime \left( \frac{1}{3} \right) - 2
- \frac{32}{81} \pi^2 \right] T_F \Nf 
+ \left[ \frac{3}{8} + \frac{23}{162} \pi^2 - \frac{23}{108} \psi^\prime 
\left( \frac{1}{3} \right) \right] C_A \right] a \nonumber \\
&& +~ O(a^2) \nonumber \\ 
\tilde{\Sigma}^{ggg}_{(2)}(p,q,0,0) &=& 
\left[ \left[ \frac{16}{27} + \frac{128}{243} \pi^2 - \frac{64}{81}
\psi^\prime \left( \frac{1}{3} \right) \right] T_F \Nf 
+ \left[ \frac{97}{108} - \frac{67}{243} \pi^2 + \frac{67}{162} \psi^\prime 
\left( \frac{1}{3} \right) \right] C_A \right] a \nonumber \\
&& +~ O(a^2) \nonumber \\ 
\tilde{\Sigma}^{ggg}_{(3)}(p,q,0,0) &=& 
\left[ \left[ \frac{64}{243} \pi^2 - \frac{28}{27} - \frac{32}{81}
\psi^\prime \left( \frac{1}{3} \right) \right] T_F \Nf 
+ \left[ \frac{67}{54} - \frac{56}{243} \pi^2 + \frac{28}{81} \psi^\prime 
\left( \frac{1}{3} \right) \right] C_A \right] a \nonumber \\
&& +~ O(a^2) 
\end{eqnarray}
for the triple gluon vertex. We note that we have checked that these 
expressions agree with those derived in \cite{38} after converting the 
amplitudes of \cite{38} to the basis used here. The amplitudes for the 
ghost-gluon vertex are
\begin{eqnarray}
\Sigma^{ccg}_{(1)}(p,q,0,0) &=& -~ 1
+ \left[ \frac{5}{108} \pi^2 - \frac{5}{72} \psi^\prime \left( \frac{1}{3} 
\right) - \frac{1}{2} \right] C_A a ~+~ O(a^2) \nonumber \\ 
\Sigma^{ccg}_{(2)}(p,q,0,0) &=& 
\left[ \frac{5}{72} \psi^\prime \left( \frac{1}{3} \right) 
- \frac{5}{108} \pi^2 + \frac{1}{4} \right] C_A a ~+~ O(a^2)
\end{eqnarray}
and those for the quark-gluon vertex are
\begin{eqnarray}
\Sigma^{qqg}_{(1)}(p,q,0,0) &=& 1
+ \left[ \left[ \frac{13}{4} + \frac{13}{54} \pi^2 - \frac{13}{36} \psi^\prime 
\left( \frac{1}{3} \right) \right] C_A 
+ \left[ \frac{2}{9} \psi^\prime \left( \frac{1}{3} \right) - \frac{4}{27} 
\pi^2 - 2 \right] C_F \right] a \nonumber \\
&& +~ O(a^2) \nonumber \\ 
\Sigma^{qqg}_{(2)}(p,q,0,0) &=& \Sigma^{qqg}_{(5)}(p,q,0,0) \nonumber \\
&=& \left[ \left[ \frac{7}{3} + \frac{5}{27} \pi^2 - \frac{5}{16} \psi^\prime 
\left( \frac{1}{3} \right) \right] C_A 
+ \left[ \frac{4}{9} \psi^\prime \left( \frac{1}{3} \right) - \frac{8}{27} 
\pi^2 - \frac{8}{3} \right] C_F \right] a ~+~ O(a^2) \nonumber \\ 
\Sigma^{qqg}_{(3)}(p,q,0,0) &=& \Sigma^{qqg}_{(4)}(p,q,0,0) \nonumber \\
&=& \left[ \left[ \frac{5}{3} + \frac{2}{27} \pi^2 - \frac{1}{9} \psi^\prime 
\left( \frac{1}{3} \right) \right] C_A 
- \frac{4}{3} C_F \right] a ~+~ O(a^2) \nonumber \\ 
\Sigma^{qqg}_{(6)}(p,q,0,0) &=&
\left[ \left[ \frac{11}{27} \pi^2 - \frac{11}{18} \psi^\prime 
\left( \frac{1}{3} \right) \right] C_A 
+ \left[ \frac{4}{9} \psi^\prime \left( \frac{1}{3} \right) - \frac{8}{27} 
\pi^2 \right] C_F \right] a ~+~ O(a^2) ~. 
\end{eqnarray}
The numerical evaluation of these as well as all the two loop corrections were
given in \cite{38}.

\sect{Expansions for various master integrals.}

In this section we present the expansions for several master integrals in 
powers of $1/\mu^2$. These were established by the methods of \cite{31} which
were discussed in section 3. First, we have for one mass scale 
\begin{eqnarray}
I(i\sqrt{C_A}\gamma^2,0,0) &=&
\left[ \frac{4\pi^2}{9} - \frac{2}{3} \psi^\prime \left( \frac{1}{3} \right) 
\right] \frac{1}{\mu^2} ~-~ \left[ \frac{\pi}{2} + i - \frac{i}{2} 
\ln \left[ \frac{C_A\gamma^4}{\mu^4} \right] \right] 
\frac{\sqrt{C_A}\gamma^2}{\mu^4} \nonumber \\
&& - \left[ \frac{3}{4} + \frac{1}{4} 
\ln \left[ \frac{C_A\gamma^4}{\mu^4} \right]
+ \frac{\pi i}{4} \right] \frac{C_A\gamma^4}{\mu^6} 
+ O \left( \frac{\gamma^6}{\mu^8} \right) ~.
\end{eqnarray}
For two non-zero masses we have 
\begin{eqnarray}
I(i\sqrt{C_A}\gamma^2,i\sqrt{C_A}\gamma^2,0) &=&
\left[ \frac{4\pi^2}{9} - \frac{2}{3} \psi^\prime \left( \frac{1}{3} \right) 
\right] \frac{1}{\mu^2} 
- \left[ \pi + 2 i - i 
\ln \left[ \frac{C_A\gamma^4}{\mu^4} \right] \right] 
\frac{\sqrt{C_A}\gamma^2}{\mu^4} \nonumber \\
&& - \left[ \frac{1}{2} + \frac{3}{2} 
\ln \left[ \frac{C_A\gamma^4}{\mu^4} \right]
+ \frac{3\pi i}{2} \right] \frac{C_A\gamma^4}{\mu^6} 
+ O \left( \frac{\gamma^6}{\mu^8} \right)
\end{eqnarray}
and
\begin{eqnarray}
I(i\sqrt{C_A}\gamma^2,-i\sqrt{C_A}\gamma^2,0) &=&
\left[ \frac{4\pi^2}{9} - \frac{2}{3} \psi^\prime \left( \frac{1}{3} \right) 
\right] \frac{1}{\mu^2} - \pi \frac{\sqrt{C_A}\gamma^2}{\mu^4} \nonumber \\
&& + \left[ \frac{1}{2} \ln \left[ \frac{C_A\gamma^4}{\mu^4} \right]
- \frac{5}{2} \right] \frac{C_A\gamma^4}{\mu^6} 
+ O \left( \frac{\gamma^6}{\mu^8} \right)
\end{eqnarray}
which is real as expected. In the pure Gribov-Zwanziger case the main cases 
with three non-zero entries are
\begin{eqnarray}
I(i\sqrt{C_A}\gamma^2,i\sqrt{C_A}\gamma^2,i\sqrt{C_A}\gamma^2) &=&
\left[ \frac{4\pi^2}{9} - \frac{2}{3} \psi^\prime \left( \frac{1}{3} \right) 
\right] \frac{1}{\mu^2} \nonumber \\
&& - \left[ \frac{3\pi}{2} + 3 i - \frac{3i}{2} 
\ln \left[ \frac{C_A\gamma^4}{\mu^4} \right] \right] 
\frac{\sqrt{C_A}\gamma^2}{\mu^4} \nonumber \\
&& + \left[ \frac{3}{4} - \frac{15}{4} 
\ln \left[ \frac{C_A\gamma^4}{\mu^4} \right]
- \frac{15\pi i}{4} \right] \frac{C_A\gamma^4}{\mu^6} 
+ O \left( \frac{\gamma^6}{\mu^8} \right)
\end{eqnarray}
and
\begin{eqnarray}
I(i\sqrt{C_A}\gamma^2,i\sqrt{C_A}\gamma^2,-i\sqrt{C_A}\gamma^2) &=&
\left[ \frac{4\pi^2}{9} - \frac{2}{3} \psi^\prime \left( \frac{1}{3} \right) 
\right] \frac{1}{\mu^2} \nonumber \\
&& - \left[ \frac{3\pi}{2} + i - \frac{i}{2} 
\ln \left[ \frac{C_A\gamma^4}{\mu^4} \right] \right] 
\frac{\sqrt{C_A}\gamma^2}{\mu^4} \nonumber \\
&& + \left[ \frac{1}{4} \ln \left[ \frac{C_A\gamma^4}{\mu^4} \right]
- \frac{13}{4} - \frac{5\pi i}{4} \right] \frac{C_A\gamma^4}{\mu^6}
+ O \left( \frac{\gamma^6}{\mu^8} \right) . ~~~~
\end{eqnarray}
For either the ${\cal Q}$ or ${\cal R}$ solutions we have
\begin{eqnarray}
I(\muIp^2,\muIm^2,\muI^2) &=&
\left[ \frac{4\pi^2}{9} - \frac{2}{3} \psi^\prime \left( \frac{1}{3} \right) 
\right] \frac{1}{\mu^2} \nonumber \\
&& + \left[ \muIp^2 \left[ \ln \left[ \frac{\muIp^2}{\mu^2} \right] - 1 \right]
+ \muIm^2 \left[ \ln \left[ \frac{\muIm^2}{\mu^2} \right] - 1 \right]
+ \muI^2 \left[ \ln \left[ \frac{\muI^2}{\mu^2} \right] - 1 \right] \right]
\frac{1}{\mu^4} \nonumber \\
&& + \left[ 
\muIp^4 \left[ \frac{1}{2} \ln \left[ \frac{\muIp^2}{\mu^2} \right] 
+ \frac{3}{4} \right]
+ \muIm^4 \left[ \frac{1}{2} \ln \left[ \frac{\muIm^2}{\mu^2} \right] 
+ \frac{3}{4} \right] \right. \nonumber \\
&& \left. ~~~
+ \muI^4 \left[ \frac{1}{2} \ln \left[ \frac{\muI^2}{\mu^2} \right] 
+ \frac{3}{4} \right] 
+ \muIp^2 \muIm^2 \left[ \ln \left[ \frac{\muIp^2}{\mu^2} \right] 
+ \ln \left[ \frac{\muIm^2}{\mu^2} \right] 
- 1 \right] \right. \nonumber \\
&& \left. ~~~
+ \muI^2 \muIp^2 \left[ \ln \left[ \frac{\muI^2}{\mu^2} \right] 
+ \ln \left[ \frac{\muIp^2}{\mu^2} \right] 
- 1 \right] \right. \nonumber \\
&& \left. ~~~
+ \muI^2 \muIm^2 \left[ \ln \left[ \frac{\muI^2}{\mu^2} \right] 
+ \ln \left[ \frac{\muIm^2}{\mu^2} \right] 
- 1 \right]
\right] \frac{1}{\mu^6} + O \left( \frac{\gamma^6}{\mu^8} \right)
\end{eqnarray}
for the case of three distinct non-zero masses. In each case the order symbols
are intended to reflect the power of $\mu$ and the numerator factor therein is
merely to have the correct overall dimensionful dependence. We have also given
the expansion out to powers beyond that which we have indicated we are 
interested in for the overall vertex functions. This is because in the 
rearrangement of the numerator scalar products in the original integrals one 
can be left with terms such as $p^2$ and $pq$ which are proportional to 
$\mu^2$. Hence, terms beyond the dimension four ones we are interested in for 
the overall vertex function need to be retained in the above examples and the 
expansion of the other basic master integrals.

\end{document}